\documentclass{aa}

\usepackage{graphicx}
\usepackage{txfonts}
\usepackage{subfigure}
\usepackage{longtable}
\usepackage{hyperref}
\begin{document}

\title{Quantifying the stellar ages of dynamically separated bulges and disks of CALIFA spiral galaxies}

\titlerunning{The stellar ages of bulges and disks of CALIFA spiral galaxies}

\author
{Yunpeng Jin\inst{1}\thanks{E-mail: jyp199333@163.com}
\and Ling Zhu\inst{1}\thanks{Corr author: lzhu@shao.ac.cn}
\and Stefano Zibetti\inst{2}
\and Luca Costantin\inst{3,4}
\and Glenn van de Ven\inst{5}
\and Shude Mao\inst{6,7}}

\institute
{Shanghai Astronomical Observatory, Chinese Academy of Sciences, 80 Nandan Road, Shanghai 200030, China
\and INAF-Osservatorio Astrofisico di Arcetri, Largo Enrico Fermi 5, I-50125 Firenze, Italy
\and Centro de Astrobiolog\'ia (CSIC-INTA), Ctra de Ajalvir km 4, Torrej\'on de Ardoz, 28850, Madrid, Spain
\and INAF-Osservatorio Astronomico di Brera, Via Brera 28, 20121, Milano, Italy
\and Department of Astrophysics, University of Vienna, Türkenschanzstraße 17, 1180 Wien, Austria
\and Department of Astronomy, Tsinghua University, Beĳing 100084, China
\and National Astronomical Observatories, Chinese Academy of Sciences, 20A Datun Road, Chaoyang District, Beĳing 100101, China}
\date{Received; accepted}

\abstract
{We employ a recently developed population-orbit superposition technique to simultaneously fit the stellar kinematic and age maps of 82 CALIFA spiral galaxies and obtain the ages of stars in different dynamical structures. We first evaluated the capabilities of this method on CALIFA-like mock data created from the Auriga simulations. The recovered mean ages of dynamically cold, warm, and hot components match the true values well, with an observational error of up to $20\%$ in the mock age maps. For CALIFA spiral galaxies, we find that the stellar ages of the cold, warm, and hot components all increase with the stellar mass of the galaxies, from $\overline{t_{\rm cold}}\sim2.2$ Gyr, $\overline{t_{\rm warm}}\sim2.3$ Gyr, and $\overline{t_{\rm hot}}\sim2.6$ Gyr for galaxies with stellar mass $M_*<10^{10}\,\rm M_{\odot}$, to $\overline{t_{\rm cold}}\sim4.0$ Gyr, $\overline{t_{\rm warm}}\sim5.1$ Gyr, and $\overline{t_{\rm hot}}\sim5.9$ Gyr for galaxies with $M_*>10^{11}\,\rm M_{\odot}$. About $80\%$ of the galaxies in our sample have $t_{\rm hot}>t_{\rm cold}$, and the mean values of $t_{\rm hot}-t_{\rm cold}$ also increase with stellar mass, from $0.7_{-0.2}^{+0.6}$ Gyr in low-mass galaxies ($10^{8.9}\,\rm M_{\odot}<M_*\le10^{10.5}\,\rm M_{\odot}$) to $1.7_{-0.2}^{+0.7}$ Gyr in high-mass galaxies ($10^{10.5}\,\rm M_{\odot}<M_*<10^{11.3}\,\rm M_{\odot}$). The stellar age is younger in disks than in bulges, on average. This suggests that either the disks formed later and/or that they experienced a more prolonged and extensive period of star formation. Lower-mass spiral galaxies have younger bulges and younger disks, while higher-mass spiral galaxies generally have older bulges, and their disks span a wide range of ages. This is consistent with the scenario in which the bulges in more massive spirals formed earlier than those in less massive spirals.}

\keywords{galaxies: spiral -- galaxies: structure -- galaxies: kinematics and dynamics -- galaxies: evolution -- galaxies: fundamental parameters}

\maketitle
\begin{nolinenumbers}

\section{Introduction}
\label{sec1}
Our knowledge of galaxy structures evolves with the development of the instruments that are used in galaxy observations. The preliminary understanding of external galaxies, such as the morphology and colour, directly comes from photometric images of galaxies. Galaxies are classified into different Hubble types based on the presence of a disk, a bulge, and/or bar structures \citep{Hubble1926}. Since the 1970s, numerous single-band surveys from the ultraviolet to the far-infrared have been carried out. They generally tell us that typical elliptical galaxies (and bulges in spiral galaxies) tend to be red, gas poor, and have no ongoing star formations, whereas the disks of spiral galaxies are usually blue, gas rich, and have distinct star formation regions \citep{Kennicutt1998}.\\

Since the late 1990s, large galaxy surveys such as the Sloan Digital Sky Survey (SDSS; \citealp{York2000}) provided multi-band images for millions of nearby galaxies. New methods such as the spectral energy distribution fitting (SED fitting; e.g. \citealp{Bolzonella2000,Walcher2011}) are used to derive the stellar age across the image plane for each galaxy. Low-mass galaxies are found to be typically young, while massive galaxies are usually old (e.g. \citealp{Kauffmann2003,Gallazzi2005}). Most spiral galaxies have clear negative age gradients (e.g. \citealp{Bell2000,MacArthur2004}), which indicates that the stellar populations of the bulge are older than those in the disk, while the age gradients in elliptical galaxies are usually mild or even negligible (e.g. \citealp{Trager2000,Mehlert2003}). It has also been found that galaxies with higher stellar velocity dispersion tend to be older (e.g. \citealp{Thomas2005,Bernardi2006}).\\

It is widely thought that bulges and disks in galaxies should be formed through different physical processes. To analyse them separately, photometric bulge-disk decomposition methods are widely used, which usually assume that the surface brightness of the bulge follows a radial profile with S\'ersic index $n\gtrsim2$ \citep{Sersic1968}, while the surface brightness of the disk follows an exponential profile with index $n=1$ (e.g. \citealp{Peng2002,Peng2010,deSouza2004,MendezAbreu2008}). The S\'ersic index derived from the photometric decomposition is widely used to separate the classical bulges that formed via major mergers and the pseudobulges, which formed via disk instability (e.g. \citealp{Fisher2008,Weinzirl2009,Gadotti2009}). However, it may not be an appropriate diagnostic for a distinction between bulges (e.g. \citealp{Graham2008,Costantin2018,Gao2022}). Recent studies have shown that the surface brightness profile of the disk may not be exactly exponential, but down-bending or up-bending in the inner or outer regions (e.g. \citealp{MendezAbreu2017,Breda2020}), which challenges the basic assumption taken in photometric decomposition. In addition, photometric decomposition provides limited information on the stellar populations of bulges and disks separately.\\

In recent years, Integral Field Spectroscopy (IFS) surveys such as SAURON \citep{Bacon2001}, $\rm ATLAS^{3D}$ \citep{Cappellari2011}, CALIFA \citep{Sanchez2012}, SAMI \citep{Bryant2015}, MaNGA \citep{Bundy2015}, and multiple MUSE programmes \citep{Bacon2017} have observed thousands of nearby galaxies. IFS observations allow us to investigate the spatially resolved stellar population distribution of a galaxy (e.g. \citealp{McDermid2015,GonzalezDelgado2015,Goddard2017,Li2018,Bernardi2019}). The traditional photometric decomposition was then updated to a spectrophotometric decomposition method \citep{MendezAbreu2019}, which fits the IFS data cube and obtains the stellar populations of photometrically decomposed bulges and disks separately. Subsequent research suggested that bulges in spiral galaxies form first and have not evolved much through cosmic time, and the disks are built up later around the bulges, but do not affect them significantly (e.g. \citealp{MendezAbreu2021}). Multiple spectroscopic decomposition attempts have been made that directly separated the bulge and disk by fitting the spectra (e.g. \citealp{Johnston2012,Johnston2014,Tabor2017,Tabor2019,Oh2020,Pak2021}) based on the full spectral fitting software pPXF \citep{Cappellari2004,Cappellari2017}. These works in general found that the bulges in spiral galaxies are older, more metal rich, and rotate more slowly than the disks, while the differences in age and metallicity between bulges and disks become small or even negligible in early-type galaxies. Recent studies of spiral galaxies at intermediate redshift $0.14<z\le1$ based on SHARDS \citep{PerezGonzalez2013} and HST/CANDELS \citep{Koekemoer2011,Grogin2011} data also found that the bulges in spiral galaxies are usually older than the disks, and their age difference increases with galaxy stellar mass, which was explained by two formation epochs of bulges in different galaxies \citep{Costantin2021,Costantin2022}.\\

Dynamical decomposition is an alternative way to distinguish between different galaxy structures, such as bulges and disks, and it is commonly used in simulations (e.g. \citealp{Correa2017,Pillepich2019,RodriguezGomez2019,Du2019,Du2020,Pulsoni2020}). This decomposition also becomes possible for real galaxies based on Schwarzschild's orbit-superposition method \citep{Schwarzschild1979,Schwarzschild1982,Schwarzschild1993} by fitting the stellar luminosity distribution and kinematic data from IFS observations. Several implementations of Schwarzschild's method are currently in use (e.g. \citealp{vdB2008,Vasiliev2020,Neureiter2021}). The triaxial Schwarzschild model developed by \citet{vdB2008} was extensively used to analyse large samples of nearby galaxies from IFS observations to obtain the mass distributions and intrinsic shapes (e.g. \citealp{Jin2019,Jin2020,Santucci2022,Pilawa2022}), internal orbital structures (e.g. \citealp{Zhu2018a,Zhu2018b,Zhu2018c,Jin2019,Jin2020,Santucci2022}), and black hole masses (e.g. \citealp{Quenneville2021,Quenneville2022,Pilawa2022}). This triaxial model has recently been modified to explicitly include a bar \citep{Tahmasebzadeh2021,Tahmasebzadeh2022}. The orbit-superposition method allows us to obtain the probability density distributions of stellar orbits in a galaxy and to carry out an orbital decomposition. Studies of CALIFA galaxies revealed that dynamically hot orbits usually occupy the inner bulge regions, while dynamically cold orbits are well correlated with the photometrically decomposed disk fraction \citep{Zhu2018a,Zhu2018c}. The hot orbit fraction $f_{\rm hot}$ decreases with stellar mass for galaxies with $M_*<10^{10}\,\rm M_{\odot}$ and increases with stellar mass for galaxies with $M_*>10^{10}\,\rm M_{\odot}$, while the cold orbit fraction $f_{\rm cold}$ increases with stellar mass below $10^{10}\,\rm M_{\odot}$ and decreases with stellar mass above $10^{10}\,\rm M_{\odot}$ \citep{Zhu2018b}. Similar mass dependences of $f_{\rm hot}$ and $f_{\rm cold}$ were also found in \citet{Jin2020} for MaNGA early-type galaxies. These results enabled direct and statistical comparison of galactic structures to cosmological simulations \citep{Xu2019}.\\

A population-orbit superposition method has further been developed based on the Schwarzschild method by tagging the stellar orbits with ages and metallicities, allowing us to obtain the stellar populations of different dynamical structures. This method was validated by \citet{Zhu2020} using MUSE-like mock data created from the Auriga simulations \citep{Grand2017}. It has been applied to 23 early-type galaxies from the Fornax3D project \citep{Sarzi2018}, leading to several new analyses of the internal galaxy structures, including the deviation of the intrinsic age-velocity dispersion profile in disks of external galaxies \citep{Poci2019,Poci2021}, weighting and timing ancient massive mergers in external galaxies using a so-called hot inner stellar halo structure \citep{Zhu2022a,Zhu2022b}, and probing the cluster environmental effects on the formation of cold galaxy disks \citep{Ding2023}.\\

In this paper, we validate the population-orbit superposition method with CALIFA-like mock data and apply it to a sample of 82 CALIFA spiral galaxies. With the stellar orbit distributions already derived from \citet{Zhu2018b} via the standard Schwarzschild method, we further obtain the stellar ages of different dynamical components. In Section 2 we introduce the population-orbit superposition method. In Section 3 we validate the method with CALIFA-like mock data, and in Section 4 we present the statistical analysis for CALIFA spiral galaxies. We discuss our findings in Section 5, and we conclude in Section 6.\\

\section{The population-orbit superposition method}
\label{sec2}
We took a two-step process to create a population-orbit superposition model for a galaxy, as proposed by \citet{Zhu2020}: We first created a standard Schwarzschild's orbit-superposition model by fitting the stellar luminosity distribution and kinematic data, and searched for the best-fitting model by exploring the free parameter space. Then we tagged the stellar orbits of the best-fitting model with stellar populations. In this paper, we only tagged stellar orbits with stellar ages, but not with metallicities.\\

\begin{figure}
\begin{centering}
    \includegraphics[width=8.5cm]{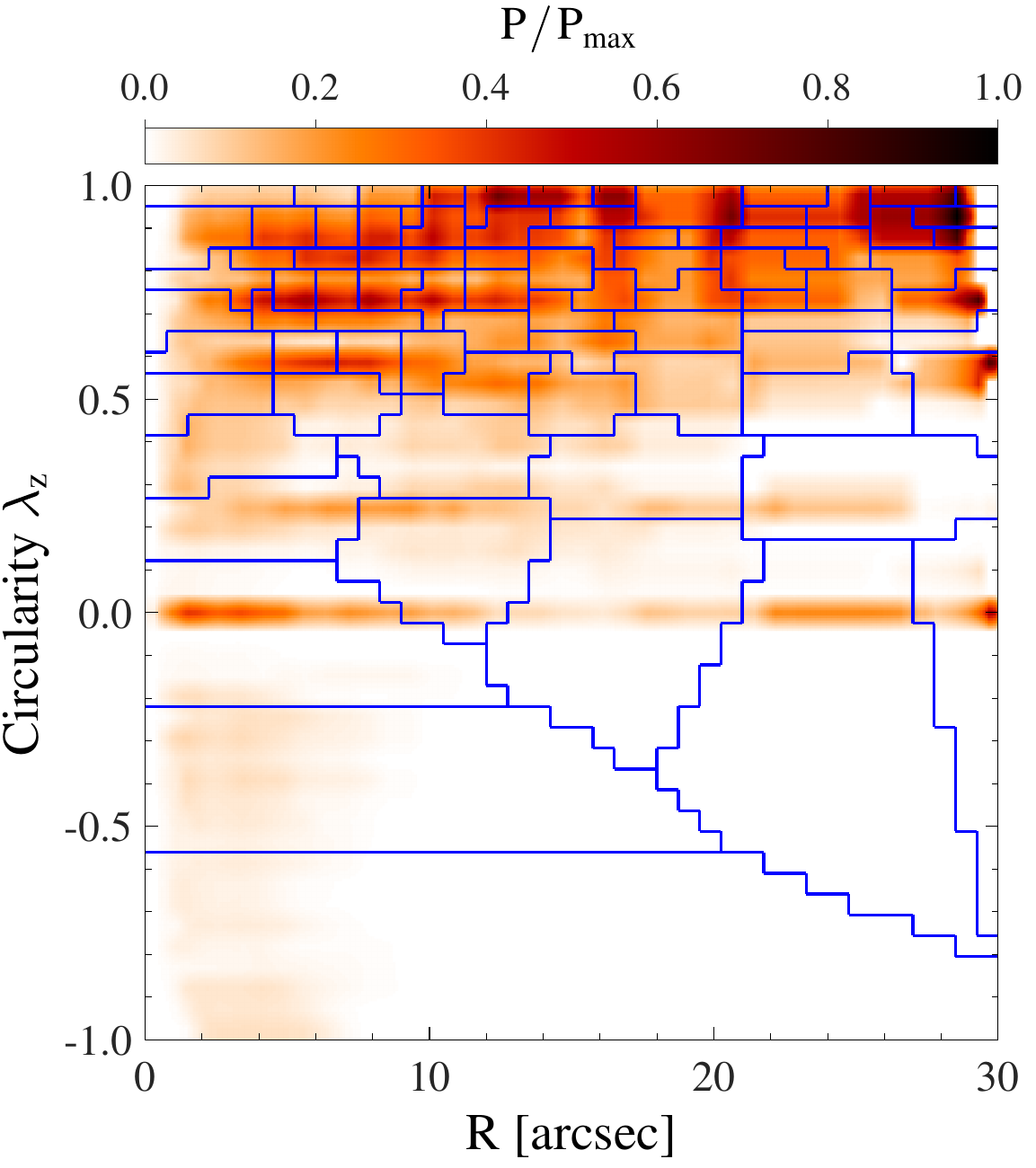}
    \caption{{\bf Probability density distributions of stellar orbits $P(\lambda_z,r)$ in the phase space of time-averaged radius $r$ vs circularity $\lambda_z$ for the best-fitting model of a typical simulated spiral galaxy.} The normalized probability $P/P_{\rm max}$ is indicated by the colour bar. We adopted the Voronoi-binning scheme in the $\lambda_z$-$r$ plane to make roughly equal-weight orbit bundles. Their boundaries are represented by the blue lines.}
    \label{fig1}
\end{centering}
\end{figure}
\begin{figure*}
\begin{centering}
    \includegraphics[width=17.8cm]{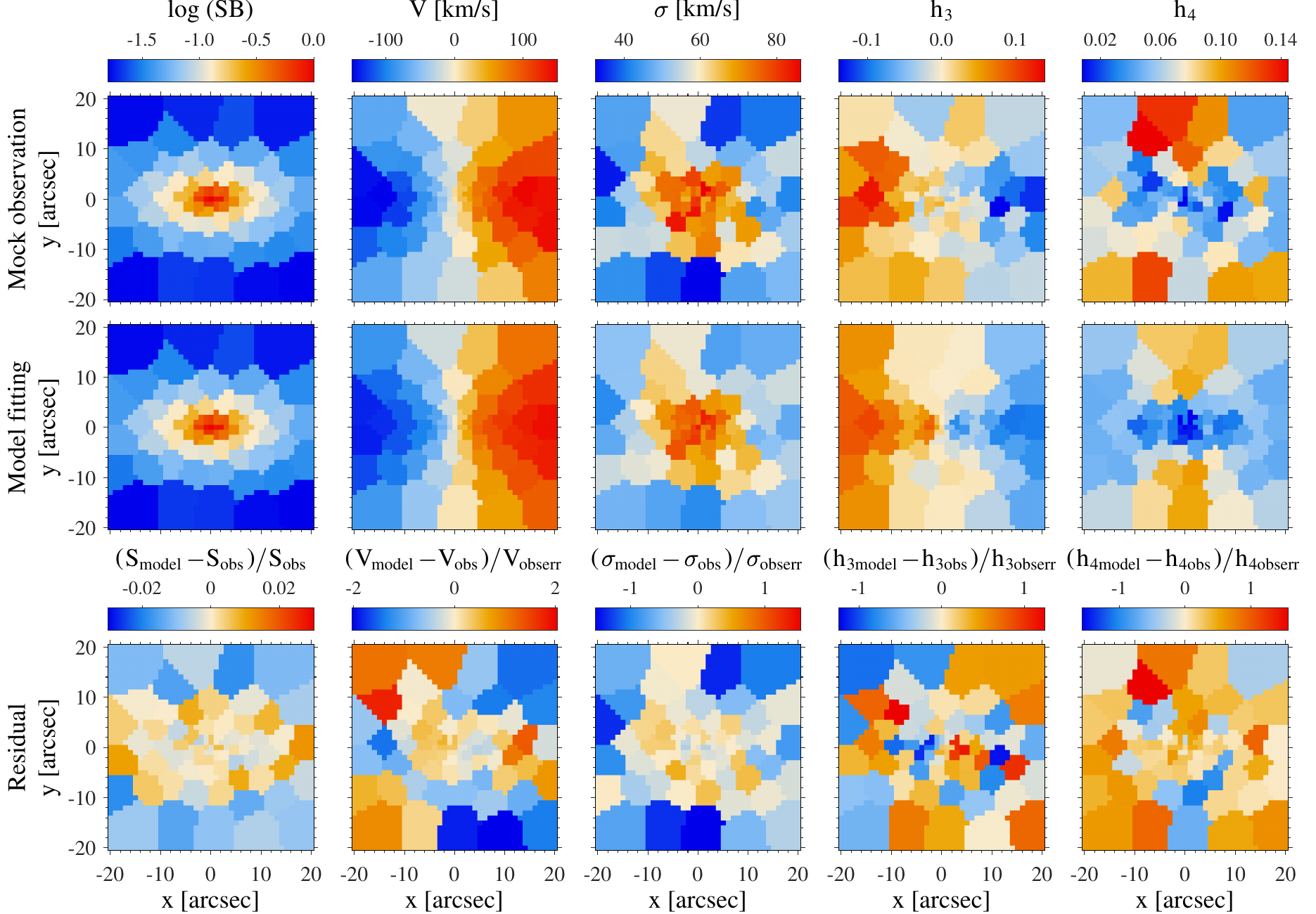}
    \caption{{\bf Mock surface brightness, stellar kinematic data, and best-fitting model for Au-6 with an inclination angle $\bf\vartheta=60^{\circ}$}. From left to right: Logarithmic normalised surface brightness, mean velocity $V$, velocity dispersion $\sigma$, skewness $h_3$, and kurtosis $h_4$. From top to bottom: Mock observations, model fittings, and residuals. The residuals represent the relative deviations of the surface brightness and the standardised residuals of the kinematics.}
    \label{fig2}
\end{centering}
\end{figure*}

\subsection{Creating an orbit-superposition model}
\label{sec2.1}
This paper follows the same approach as \citet{vdB2008} and \citet{Zhu2018b}. We took three steps to create an orbit-superposition Schwarzschild model for a galaxy: (1) We constructed the gravitational potential, (2) sampled and integrated orbits, and (3) solved the orbit weights by fitting the observational data, which included the 2D surface brightness, the 3D luminosity density distribution, and kinematic data.\\

The gravitational potential of the galaxy was generated by combining a triaxial stellar component, a dark matter halo, and a central supermassive black hole that was treated as a point source. To calculate the stellar potential, the galaxy surface brightness was first fitted by the multi-Gaussian expansion (MGE) formalism \citep{Emsellem1994,Cappellari2002}. The 3D luminosity density was then deprojected from the 2D MGE by setting the viewing angles $(\theta,\phi,\psi)$. By multiplying the 3D luminosity density by the mass-to-light ratio $M_*/L$, the stellar mass distribution was calculated, and subsequently the stellar gravitational potential was derived by solving Poisson's equation. The dark matter potential was described by a spherical Navarro–Frenk–White (NFW) profile \citep{Navarro1996}, which had two free parameters: the enclosed halo mass $M_{\rm 200}$ within the virial radius, and the dark matter concentration $c$. The viewing angles $(\theta,\phi,\psi)$ were transformed into the intrinsic triaxial shapes of the galaxy: the axis ratios $p=b/a$, $q=c/a$, and the compression factor $u$. In practice, to reduce degeneracy, the concentration $c$ was fixed following the $M_{\rm 200}-c$ relation from \citet{Dutton2014}, and the compression factor was set to $u=0.9999$. Thus, there were four free parameters in total: $M_*/L$, $p$, $q$, and $M_{\rm 200}$.\\

The orbits were sampled from three integrals of motion that define a triaxial system: energy $E$, second integral $I_2$, and third integral $I_3$. The orbit libraries include a typical library of orbits sampled from the $x-z$ plane, a library of counter-rotating orbits, and an additional library of box orbits sampled from the equipotential plane. When modelling CALIFA galaxies \citep{Zhu2018b,Zhu2018c}, a combination of $21\times10\times7$ intervals in ($E, I_2, I_3$) was used to generate the initial orbit conditions in each orbit library. Then, by slightly perturbing the initial conditions, each orbit was dithered to $5^3$ orbits  to form an orbit bundle. Thus, there are $3\times21\times10\times7=4410$ orbit bundles in total. \\

The orbit weights were then solved using the non-negative least-squares (NNLS; \citealp{Lawson1974}) implementation, taking both stellar luminosity distribution and kinematic data as constraints. The $\chi_{\rm err}^2$ that needs to be minimised by NNLS is expressed as
\begin{equation}
    \chi_{\rm err}^2=\chi_{\rm lum}^2+\chi_{\rm kin}^2.
\end{equation}
For the luminosity distribution, we used both the 2D surface brightness and 3D luminosity density as constraints. The 2D surface brightness was stored in the same apertures as the kinematic data, with $S_i$ representing the surface brightness in the $i$th aperture. The 3D space was divided into 360 cells, with $\gamma_m$ representing the luminosity of the $m$th cell. We set the relative errors of $S_i$ to be $1\%$ and $\gamma_m$ to be $2\%$. For the kinematic data, we fit the Gauss–Hermite coefficients \citep{Gerhard1993,vdMarel1993,Rix1997} of the line-of-sight velocity distribution. The Gauss–Hermite coefficients and their errors $h_1$, $h_2$, $\Delta h_1$, and $\Delta h_2$ were converted from the observational data $V$, $\sigma$, $\Delta V$, and $\Delta \sigma$. Thus, we have
\begin{equation}
    \chi_{\rm lum}^2=\sum_{\rm i=1}^{N_{\rm obs}}\left(\frac{S_i^*-S_i}{0.01S_i}\right)^2+\sum_{\rm m=1}^{M}\left(\frac{\gamma_m^*-\gamma_m}{0.02\gamma_m}\right)^2,
\end{equation}
\begin{equation}
\begin{split}
    &\chi_{\rm kin}^2=\sum_{\rm i=1}^{N_{\rm obs}}\left[ \left(\frac{h^*_{1i}-h_{1i}}{\Delta h_{1i}}\right)^2+\left(\frac{h^*_{2i}-h_{2i}}{\Delta h_{2i}} \right)^2+\right.\\
    &\ \ \ \ \ \ \ \ \ \ \ \ \ \ \ \ \ \left.\left(\frac{h^*_{3i}-h_{3i}}{\Delta h_{3i}} \right)^2+\left(\frac{h^*_{4i}-h_{4i}}{\Delta h_{4i}} \right)^2\right],\\
\end{split}
\end{equation}
where $N_{\rm obs}$ is the number of 2D apertures, and $M$ is the number of 3D cells. Variables marked with an asterisk indicate the model fittings, those with an upward-pointing triangle represent the errors of the input data, and those without a marker denote the input data.\\

After solving the orbit weights, we calculated the difference between model-recovered and observed kinematic maps directly,
\begin{equation}
\begin{split}
    &\chi^2=\sum_{\rm i=1}^{N_{\rm obs}}\left[ \left(\frac{V^*_i-V_i}{\Delta V_i}\right)^2+\left(\frac{\sigma^*_i-\sigma_i}{\Delta \sigma_i} \right)^2+\right.\\
    &\ \ \ \ \ \ \ \ \ \ \ \ \ \ \ \ \ \left.\left(\frac{h^*_{3i}-h_{3i}}{\Delta h_{3i}} \right)^2+\left(\frac{h^*_{4i}-h_{4i}}{\Delta h_{4i}} \right)^2\right],\\
\end{split}
\label{chi2kin}
\end{equation}
which, in principle, should be highly correlated with $\chi^2_{\rm kin}$ we fitted. However, if the line-of-sight velocity distribution significantly deviates from a Gaussian distribution, $\chi^2_{\rm kin}$ and $\chi^2$ may not be similar. Because $\chi^2$ directly reflects how the model matches the data, we took $\chi^2$ to represent the goodness of fit. In practice, luminosity distributions are much easier to fit than kinematic data, so that $\chi_{\rm lum}^2$ is a low value and was not taken into account in the goodness evaluation.\\

We have four free hyper-parameters ($M_*/L$, $p$, $q$, and $M_{\rm 200}$), and we took an iterative process to search for the best-fitting model. We started by making initial guesses of the parameters. After constructing the initial models, we selected the models within a relatively lower $\chi^2$ and created new models around these selected models by walking a few steps in the parameter space. We initially took large step sizes of 0.1, 0.05, 0.05, and 0.5 for $M_*/L$, $p$, $q$, $u$, and $M_{\rm 200}/M_{*}$ to avoid becoming stuck in a local minimum. We repeated this iterative process until we reached a minimum $\chi^2$ with all models around it calculated in the parameter space. Then we reduced the step size to half and repeated the process again. Finally, we determined the model with the minimum $\chi^2$, and all the models around it in the parameter space within $3\sigma$ confidence level were calculated.\\

In the end, we not only constrained the free parameters in the potential, but also obtained the best-fitting model whose stellar orbit distribution is representative of the real galaxy. Following \citet{Zhu2018b, Zhu2018c}, we characterised each orbit by its time-averaged radius $r$ and circularity $\lambda_z$. Circularity $\lambda_z$ represents the angular momentum of the orbit around the short axis normalised by the maximum of a circular orbit with the same binding energy. Therefore, we obtained the probability density distributions of stellar orbits in the phase space of $\lambda_z$ versus $r$, as illustrated in Fig.~\ref{fig1}, taking a typical simulated spiral galaxy as an example.\\

\subsection{Tagging the stellar orbits with ages}
\label{sec2.2}
After obtaining the best-fitting model in the previous step, we tagged the stellar orbits of the best-fitting model with stellar ages. We took the basic assumption that the stars on similar orbits have similar stellar populations. Following \citet{Zhu2020}, we divided the orbits in $\lambda_z$-$r$ phase space into different bundles using the Voronoi 2D binning method \citep{Cappellari2003}, ensuring that each bundle contained at least $0.5\%$ of the total orbit weight, which resulted in $N_b\sim100$ orbit bundles, as illustrated in Fig.~\ref{fig1}. We assumed that stars in the same orbit bundle are represented by a simple stellar population of age $t_k$ to be determined. Then we tagged the orbit bundles with ages, projected them onto the observing plane, and extracted a map of the stellar ages. The age $t_{\rm model}^i$ of each aperture $i$ in the observing plane was calculated by
\begin{equation}
    t_{\rm model}^i=\frac{\sum_{k=1}^{N_b} t_k f_k^i}{\sum_{k=1}^{N_b} f_k^i},
\end{equation}
where $N_b$ is the total number of orbit bundles, and $f_k^i$ is the luminosity contributed by the orbit bundle $k$ in aperture $i$. The age $t_k$ of each orbit bundle could be solved by matching the model-predicted age map with the observed data.\\

We adopted a Bayesian analysis to match the model-predicted age $t_{\rm model}^i$ with the observational age map $t_{\rm obs}^i$. To do this, we employed the Python package PyMC3\footnote{\url{https://pypi.org/project/pymc3}}, which is a probabilistic programming library that allows users to build Bayesian models and fit them with Markov chain Monte Carlo (MCMC) methods. This package allows us to approximate the posterior likelihoods, and here we adopted the Student T distribution. We began the MCMC chain with an age of $t_k^{\rm start}$, which was sampled from a bounded Gaussian distribution,
\begin{equation}
    t_k^{\rm start}\sim N(\mu_k,\sigma_k, [0\ 14]),
\end{equation}
where $\mu_k$ is the centre of the Gaussian, and $\sigma_k$ represents the dispersion. We set the physical restriction that the age value must be between 0 and 14 Gyr. The starting values of age $t_k^{\rm start}$ were initialised using the method called automatic differentiation variational inference (Advi) with 200000 draws. Then we ran 2000 steps and took the mean age of the last 500 steps as $t_k$ of each orbit bundle.\\

Because our aim is to probe the internal 3D age distribution constrained by the 2D age map with limited data quality such as CALIFA, the age of stellar orbit $t_k$ may not be fully constrained by the data, but may be affected by the starting values $t_k^{\rm start}$ in the Bayesian analysis. In the next section, we discuss the effects of choosing different $\mu_k$ and $\sigma_k$.\\

\section{Method validation with mock data}
\label{sec3}
\begin{figure}
\begin{centering}
	\includegraphics[width=8.5cm]{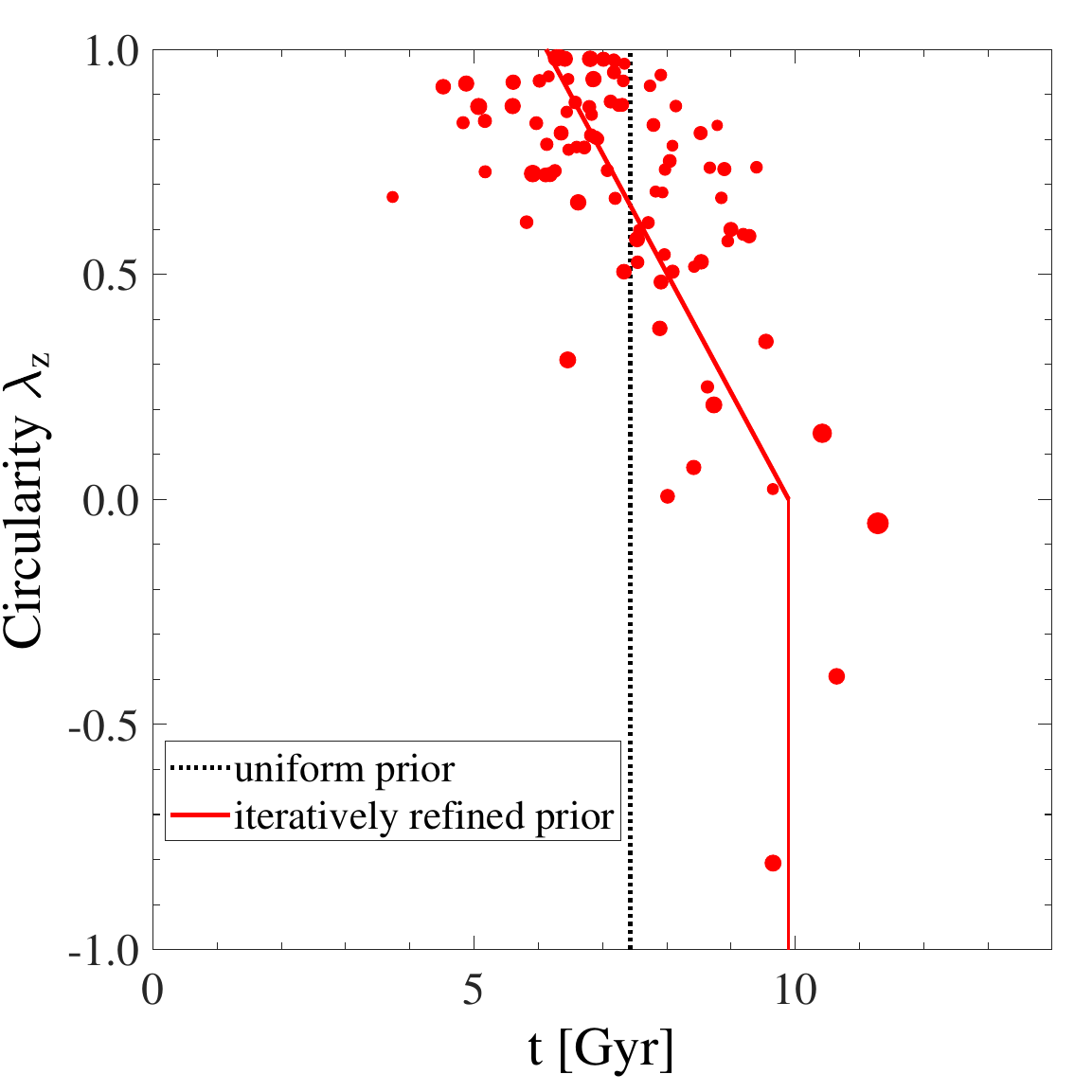}
    \caption{{\bf Correlation between stellar age $t$ and circularity $\lambda_z$ demonstrated with Au-6-60-5.} The dotted black line represents $\mu_k$ of the uniform prior we started with, while the solid red line represents $\mu_k$ of the iteratively refined prior based on the results from the previous model run. The red dots represent the final $t_k$ obtained in the last model iteration. Larger dot sizes indicate larger orbit weights of the orbit bundles.}
    \label{fig3}
\end{centering}
\end{figure}
\begin{figure}
\begin{centering}
	\includegraphics[width=8.5cm]{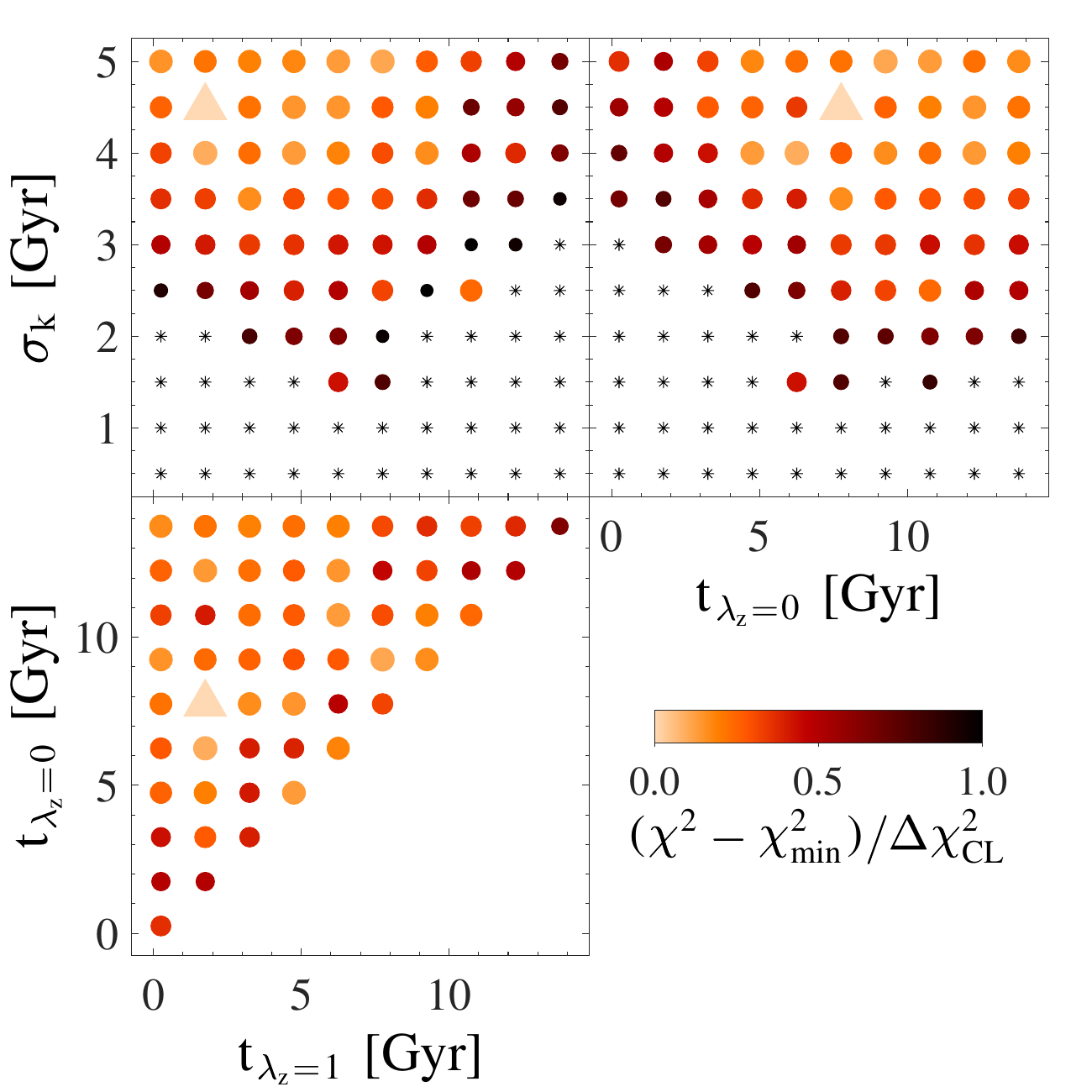}
    \caption{{\bf Parameter space of the age-circularity priors for Au-6-60-5.} We use three parameters to determine the $\lambda_z$-$t$ prior, including the age when $\lambda_z=0$, the age when $\lambda_z=1$, and the dispersion $\sigma_k$. The largest orange triangles denote the model with the minimum $\chi^2$, and the other coloured dots represent the models within the $1\sigma$ confidence level, as indicated by the colour bar. The small black asterisks represent the models outside the $1\sigma$ confidence level.}
    \label{fig4}
\end{centering}
\end{figure}
\begin{figure}
\begin{centering}
	\includegraphics[width=8.5cm]{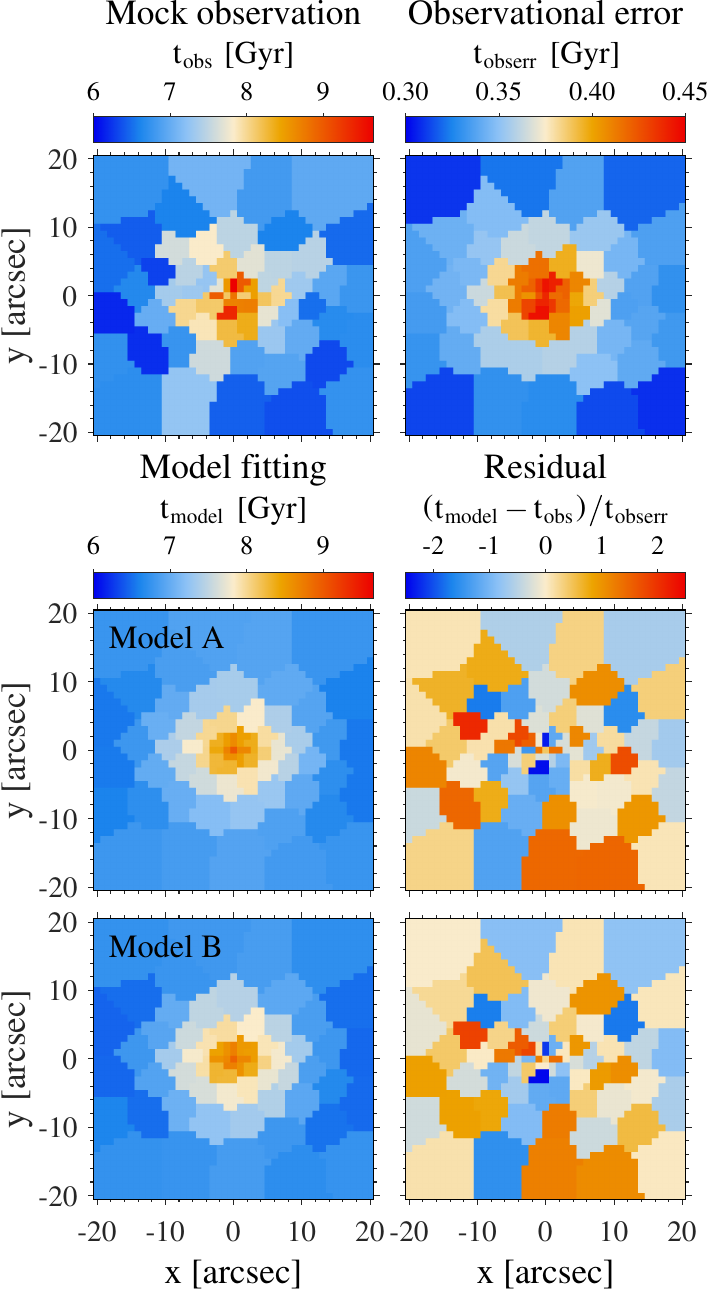}
    \caption{{\bf Mock age map and best-fitting models for Au-6-60-5}. Top panels: Mock age map $t_{\rm obs}$ and mock error map $t_{\rm obserr}$. Middle panels: Best-fitting age map $t_{\rm model}$ and standardised residuals between mock observations and model fittings $t_{\rm stdres}=(t_{\rm model}-t_{\rm obs})/t_{\rm obserr}$ derived from model A. Bottom: Similar to the middle panels, but derived from model B.}
    \label{fig5}
\end{centering}
\end{figure}

\subsection{The Auriga simulations}
\label{sec3.1}
The Auriga project is a suite of cosmological magnetohydrodynamical zoom-in simulations, which consists of 30 isolated Milky Way-mass haloes \citep{Grand2017}. The Auriga sample was selected from the dark matter-only version of the EAGLE Ref-L100N1504 simulation \citep{Schaye2015}, and the simulations were carried out with the N-body magnetohydrodynamical moving-mesh code AREPO \citep{Springel2010}. The simulations contain comprehensive physical processes of galaxy formation, such as primordial and metal-line cooling mechanism \citep{Vogelsberger2013}, a hybrid multi-phase star formation model \citep{Springel2003}, stellar feedback \citep{Marinacci2014}, AGN feedback \citep{Springel2005}, a UV background field \citep{FaucherGiguere2009}, and magnetic fields \citep{Pakmor2013}. The initial masses of baryon particles and high-resolution dark matter particles were $m_{\rm b}\sim5\times10^4\,\rm M_{\odot}$ and $m_{\rm DM}\sim3\times10^5\,\rm M_{\odot}$, respectively. We refer to \citet{Grand2017} for more details of the Auriga simulations.\\

\subsection{Mock data}
\label{sec3.2}
In this paper, we selected the three simulated galaxies Au-5, Au-6, and Au-23 from Auriga with different stellar masses $M_*$, dark matter halo masses $M_{\rm 200}$, and structures. Both Au-5 ($M_*=6.72\times10^{10}\,\rm M_{\odot}$, $M_{\rm 200}=1.19\times10^{12}\,\rm M_{\odot}$) and Au-6 ($M_*=4.75\times10^{10}\,\rm M_{\odot}$, $M_{\rm 200}=1.04\times10^{12}\,\rm M_{\odot}$) have spiral arms and a weak bar in their structures, while Au-23 ($M_*=9.02\times10^{10}\,\rm M_{\odot}$, $M_{\rm 200}=1.58\times10^{12}\,\rm M_{\odot}$) has warps and a strong bar. For each simulation, we created three versions of mock data with inclination angles ranging from close to face-on to close to edge-on: $\vartheta=40^{\circ}$, $60^{\circ}$, and $80^{\circ}$.\\

We projected each simulated galaxy onto the observational plane with a particular inclination angle $\vartheta$, placed it at a distance of 80 Mpc, and observed it as a real galaxy. We first observed it with a pixel size of 1 arcsec and calculated the stellar mass of the particles in each pixel to obtain a surface mass density map. Then we created mock stellar kinematic and age maps with a spatial coverage and resolution similar to those of the data from CALIFA survey. For each pixel, we calculated signal-to-noise ratio $S/N$ by taking the particle numbers as the signal and assuming Poisson noise. Then we performed the Voronoi-binning scheme to obtain apertures with their sizes similar to those of CALIFA galaxies, which was accomplished by setting a target $S/N=50$ with the Auriga simulation. For each aperture, we calculated the flux and the line-of-sight velocity distribution, and fit the latter with a Gauss–Hermite distribution. We thus obtained the mean velocity $V$, the velocity dispersion $\sigma$, the skewness $h_3$, and the kurtosis $h_4$. We note that the binned target $S/N=50$ is higher than the $S/N$ of CALIFA data. We further perturbed the kinematic data by adding random noise following the method in \citet{Tsatsi2015}, so that the uncertainties of the perturbed kinematic data were similar to those of the CALIFA data.\\

We took each mock data set as an observed galaxy and created the orbit-superposition model to fit the data. In Fig.~\ref{fig2} we show Au-6 with an inclination angle $\vartheta=60^{\circ}$ as an example to demonstrate the mock data, the best-fitting model, and their residuals.\\

We constructed the age maps using the same binning scheme as for the kinematic maps. We first calculated the mass-weighted mean age $t^i$ of the particles in each aperture $i$ and then added the Gaussian-distributed relative errors to the ages. Thus, the observational age map $t_{\rm obserr}^i$ and the error map $t_{\rm obserr}^i$ were written as
\begin{equation}
\begin{split}
    &t_{\rm obs}^i=t^i+t^i\times N(0,x),\\
    &t_{\rm obserr}^i=x\times t^i,\\
\end{split}
\end{equation}
where $N(0,x)$ denotes the Gaussian distribution with centre $0$ and dispersion $x$. Here, $x$ represents the relative error of the age map, and we took $x=5\%$, $10\%$, $15\%$, and $20\%$ in our analysis. By creating four versions of the age maps with different errors, we finally had $3\times3\times4=36$ mock age data sets. We used Au-6 with an inclination $\vartheta=60^{\circ}$ and an error of $5\%$ of the age map (denoted Au-6-60-5 hereafter) to illustrate our method and results.\\

\subsection{Priors of the Bayesian analysis}
\label{sec3.3}
For each mock data set, we first obtained the stellar orbit distribution by fitting the luminosity distribution and kinematic data. We then adopted a Bayesian analysis to fit the mock age map. Recent works have shown that stellar ages are strongly correlated with stellar kinematics in both real observations (e.g. \citealp{Zhu2022a}) and simulations (e.g. \citealp{Trayford2019,Bird2013,Stinson2013}). Old stars are usually dominated by random motion, while young stars tend to form a rotation-dominated disk. Therefore, we used the correlation between the stellar age and the orbit circularity to set the prior of the stellar age of the orbit.\\

\subsubsection{Single-age prior: Model A}
\label{sec3.3.1}
Following \citet{Zhu2020}, we determined the age-circularity prior by iterative model fitting. We first started the PyMC3 procedure with a uniform prior, where ($\mu_k$,$\sigma_k$) were set to the mean value and twice the standard deviation of the observational age map,
\begin{equation}
\begin{split}
    &\mu_k=\langle t_{\rm obs}^i \rangle,\\
    &\sigma_k=2\sigma(t_{\rm obs}^i).
\end{split}
\label{priorA1}
\end{equation}
Then we processed a linear fitting to the outcome $t_k(\lambda_z)$ from the previous model and took the fitting results as a new prior,
\begin{equation}
\begin{split}
    &\mu_k=t_{\lambda_z=0}+(t_{\lambda_z=1}-t_{\lambda_z=0})\lambda_z, &\rm{if}\ \lambda_z \ge 0;\\
    &\mu_k=t_{\lambda_z=0}, &\rm{if}\ \lambda_z<0;\\
    &\sigma_k=2\sigma(t_{\rm obs}^i).
\end{split}
\label{priorA2}
\end{equation}
Here, $\mu_k$ equals a constant value when $\lambda_z<0$, and follows a linear relation with the circularity $\lambda_z$ when $\lambda_z\ge0$. We reran the model with the refined prior from the previous model iteratively. The results usually converged after one or two iterations. We took the average ages of the last 500 steps in the MCMC chains from the last model iteration as our final results. In Fig.~\ref{fig3} we show how the prior $\mu_k$ is refined in a model of Au-6-60-5. The scatter between the final results and the refined prior is correlated with $\sigma_k$. We call this iterative model model A.\\

\subsubsection{Multiple age priors: Model B}
\label{sec3.3.2}
With limited data quality, we tried to improve the recovery of the stellar age distributions by introducing a revised model, model B. This model followed the same linear relation as model A, but allowed different values of $t_{\lambda_z=0}$, $t_{\lambda_z=1}$ and $\sigma_k$. We set $t_{\lambda_z=0}$ and $t_{\lambda_z=1}$ to range from 0.25 to 13.75 Gyr with an interval of 0.75 Gyr, and $\sigma_k$ to range from 0.5 to 5 Gyr with an interval of 0.5 Gyr. Because we expect the bulges to be older than the disks for most galaxies, we added an additional restriction that the priors should satisfy $t_{\lambda_z=0}\ge t_{\lambda_z=1}$. In Fig.~\ref{fig4} we show the parameter space of ($t_{\lambda_z=0}$, $t_{\lambda_z=1}$, $\sigma_k$), taking Au-6-60-5 as an example. For each set of priors, we created a population-orbit superposition model and calculated the $\chi_{\rm age}^2$ between the mock age map and the corresponding model fitting,
\begin{equation}
    \chi_{\rm age}^2=\sum_{i=1}^{N_{\rm obs}}(t_{\rm obs}^i-t_{\rm model}^i)^2,\\
\label{chi2}
\end{equation}
where $N_i$ indicates the number of apertures of age map. We used $\chi_{\rm min}^2$ to represent the minimum $\chi_{\rm age}^2$ in all sets of prior and defined the $1\sigma$ confidence level as
\begin{equation}
    \chi_{\rm age}^2-\chi_{\rm min}^2<\chi_{\rm CL}^2=\sqrt{2N_{\rm obs}}.\\
\label{CL}
\end{equation}
We averaged the age distributions of all models within the $1\sigma$ confidence level in the phase space of $\lambda_z$ versus $t$ and took this smoothed age distribution as the result from model B.\\

In Fig.~\ref{fig5} we show the best-fitting models from models A and B for Au-6-60-5. Models A and B both fit the age map well, with a relatively old inner region ($t\sim9$ Gyr) and a relatively young disk ($t\sim6$ Gyr). The standardised residuals between the mock observations and the model fittings are smaller than unity for more than half of the apertures.\\

\subsection{Test results}
\label{sec3.4}
\begin{figure*}
\begin{centering}
	\includegraphics[width=17.8cm]{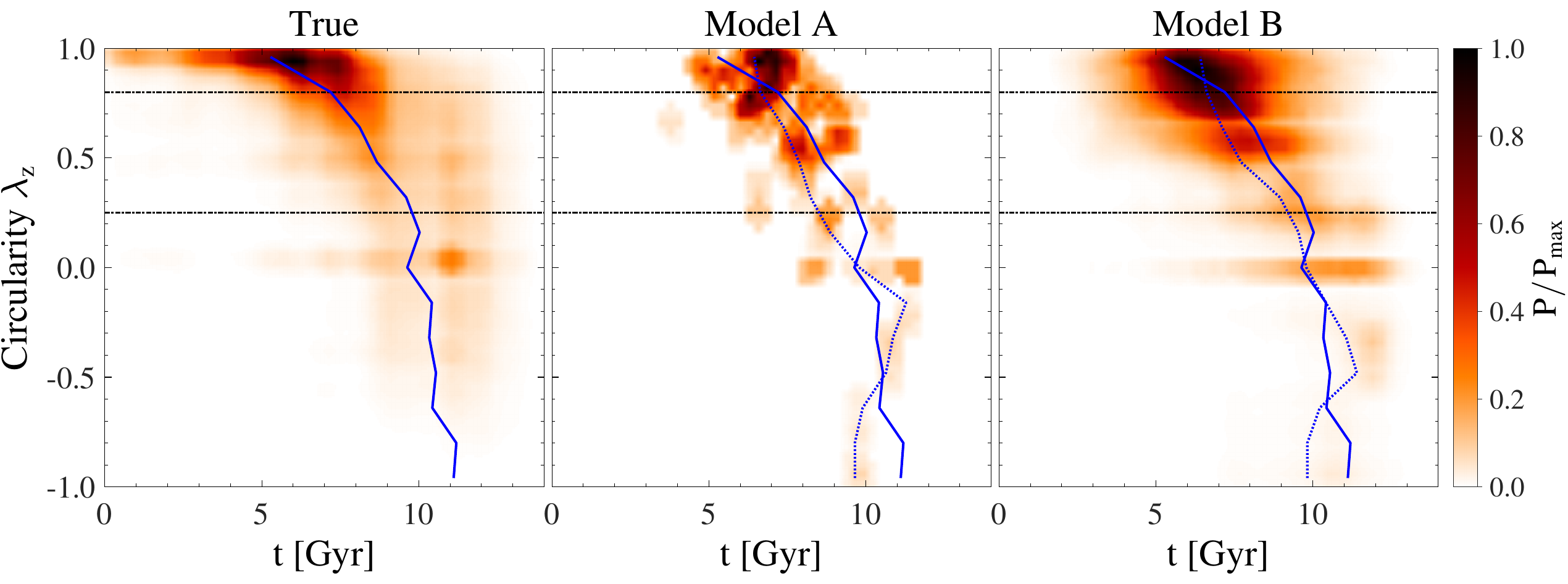}
    \caption{{\bf Probability density distributions of stellar orbits $P(\lambda_z,t)$ in the phase space of circularity $\lambda_z$ vs age $t$ for Au-6-60-5.} Left panel: True distribution directly derived from the simulation. Middle panel: Model-recovered distribution from model A. Right panel: Model-recovered distribution from model B, which is the average of all acceptable models within the $1\sigma$ confidence level. The normalised probability $P/P_{\rm max}$ is indicated by the colour bar. The horizontal dashed lines divide the cold ($\lambda_z\ge0.8$), warm ($0.25<\lambda_z<0.8$) and hot ($\lambda_z\le0.25$, including the counter-rotating orbits) components. The solid blue line in each panel represents the mean stellar age as a function of circularity from the simulations, while the dashed blue lines denote the model-recovered ages from models A and B.}
    \label{fig6}
\end{centering}
\end{figure*}

\begin{figure*}
\begin{centering}
	\includegraphics[width=17.8cm]{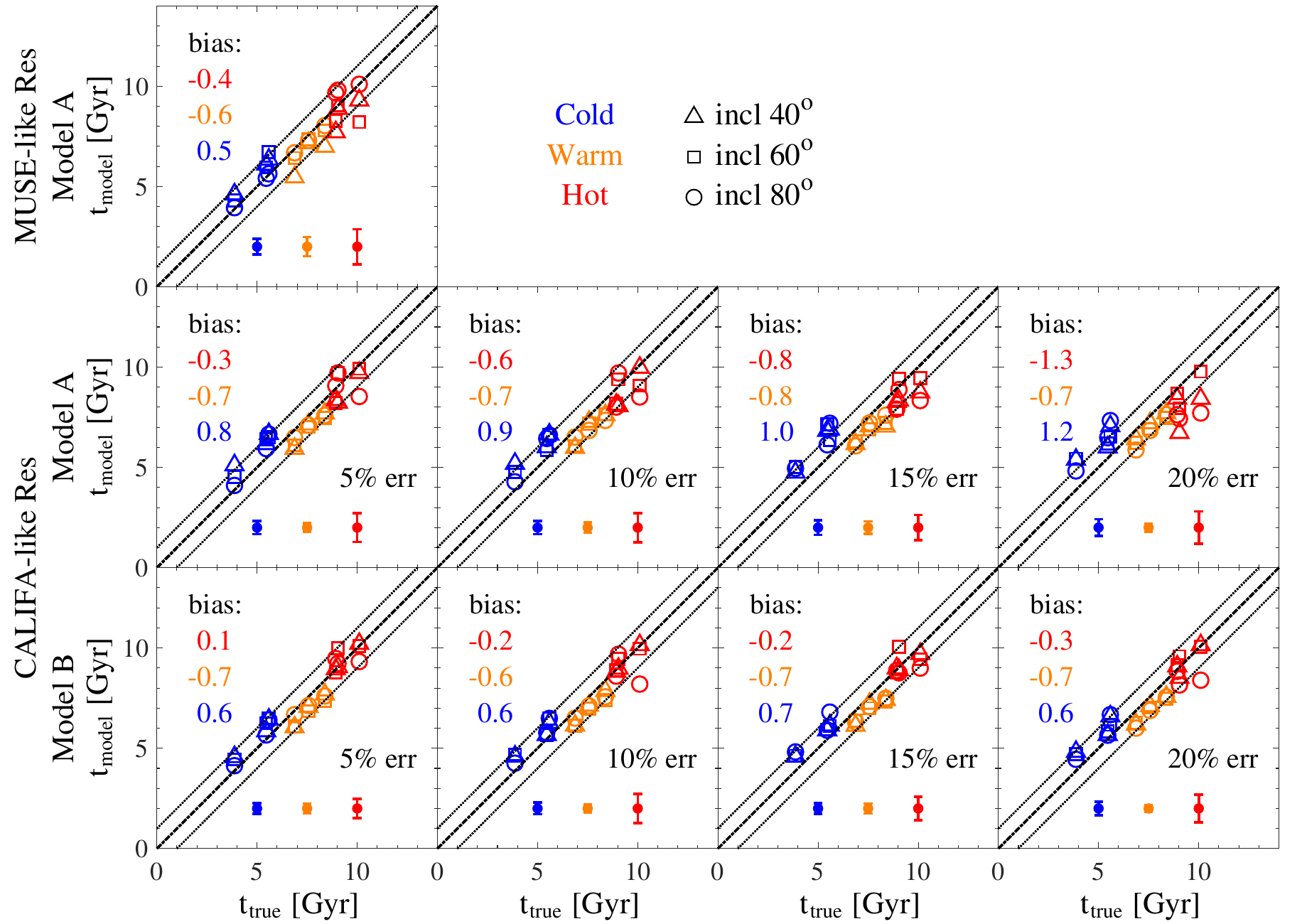}
    \caption{{\bf One-to-one comparison of the true and model-recovered age of the cold, warm, and hot orbit components for all mock data sets.} Top left panel: Results using model A and mock data with a MUSE-like spatial resolution from \citet{Zhu2020}. Middle panels: Results using model A and mock data with a CALIFA-like spatial resolution. Bottom panels: Results using model B and mock data with a CALIFA-like spatial resolution. In the middle and bottom panels, the relative observational errors equal $5\%$, $10\%$, $15\%$, and $20\%$ from left to right. The triangles, squares, and circles represent the mock data sets with inclination angles of $40^{\circ}$, $60^{\circ}$, and $80^{\circ}$. Blue, orange, and red markers denote the cold ($\lambda_z\ge0.8$), warm ($0.25<\lambda_z<0.8$), and hot ($\lambda_z\le0.25$) components. Their corresponding mean bias and statistical uncertainty are shown by the coloured texts and error bars in each panel. The dashed lines represent the equality of $t_{\rm model}$ and $t_{\rm true}$, while the dotted lines are $\pm1$ Gyr away from the dashed lines.}
    \label{fig7}
\end{centering}
\end{figure*}

In this subsection, we analyse the results obtained from our simulation tests in detail, including a comparison between models A and B. We show the orbit probability density distributions in the phase space of circularity $\lambda_z$ versus age $t$ for Au-6-60-5 in Fig.~\ref{fig6}. From left to right, the true distribution directly derived from the stellar particles in the simulation is shown, followed by the model-recovered distribution from model A and that from model B. For a fair comparison, only the stellar particles within the mock data coverage are included. Model B is smoother than model A, as model B is an average of different models, while model A is just a single model. By dividing the particles or orbits into multiple circularity bins and calculating the mean age of the particles or orbits in each bin, we find that the mean age as a function of circularity from both model A and model B generally match the true one from simulations. Despite the limited data quality, both models recover the mean age as a function of circularity well, but the detailed distribution of stellar ages is not fully recovered. For example, young stars with $t<3$ Gyr are not well represented by our models.\\

To quantify the test results, we divided each galaxy into three coarse components: cold ($\lambda_z\ge0.8$), warm ($0.25<\lambda_z<0.8$), and hot ($\lambda_z\le0.25$), as illustrated in Fig.~\ref{fig6}. The hot component here includes the counter-rotating ($\lambda_z<-0.25$) orbits as defined in \citet{Zhu2018a,Zhu2018b,Zhu2018c}. We then calculated the true and model recovered mean age of each component for all mock data sets, and we present them in Fig.~\ref{fig7}. The middle and bottom panels show our results from model A and model B, respectively, with the relative observational error ranging from $5\%$ to $20\%$. We also show the results from \citet{Zhu2020} in the top left panel. They applied model A to mock data with a MUSE-like spatial resolution.\\

When the observational error is $5\%$, model A is able to recover the mean age of each component well, similar to the results from MUSE-like mock data \citep{Zhu2020}. However, with increasing observational error, the estimations of the age of the cold and hot components ($t_{\rm cold,model}$ and $t_{\rm hot,model}$) in model A become more biased. For the mock age maps with an observational error of $5\%$, $10\%$, $15\%$, and $20\%$, the models overestimate the mean age of the cold component by $t_{\rm cold,model}-t_{\rm cold,true}=$0.8, 0.9, 1.0, and 1.2 Gyr and underestimate the mean age of the hot component by $t_{\rm hot,model}-t_{\rm hot,true}=$-0.3, -0.6, -0.8, and -1.3 Gyr, respectively. Model A, which started from a uniform prior, has a weak ability to distinguish the ages of different components when the observational error is as large as $20\%$. \\

The results of model B are shown in the bottom panels of Fig.~\ref{fig7}. The mean biases of the recovered ages are within 0.3 Gyr for the cold components and within 0.7 Gyr for the hot components, with up to $20\%$ observational error. These biases are much smaller than those of model A, and the statistical uncertainties of model B are also generally smaller. These results indicate that model B is more effective than model A, especially for data with relatively large observational errors. In both models, the ages of the cold, warm, and hot components are recovered similarly well for galaxies with different inclination angles ($40^{\circ}$, $60^{\circ}$, and $80^{\circ}$).\\

When modelling real galaxies such as the CALIFA galaxies, we used the test results with an observational error of $20\%$ to estimate the overall uncertainties of the ages of different components. We converted the absolute overall uncertainties, which included the statistical uncertainty and the mean systematic bias as shown in Fig.~\ref{fig7}, into the relative ones. Model A has a statistical uncertainty of $\sigma_{\rm stat}=11\%$, $3\%$ ,and $8\%$ and a systematic bias of $\overline{\mathcal{D}}=25\%$, $-9\%$, and $-14\%$ for the cold, warm, and hot components, respectively. Model B has a statistical uncertainty of $\sigma_{\rm stat}=8\%$, $2\%$, and $7\%$, and a systematic bias of $\overline{\mathcal{D}}=14\%$, $-9\%$, and $-3\%$.

\section{CALIFA spiral galaxies}
\label{sec4}
\subsection{CALIFA survey and dynamical models}
\label{sec4.1}
The Calar Alto Legacy Integral Field Area Survey (CALIFA; \citealp{Sanchez2012}) is an IFS survey that was designed to observe different types of galaxies representative of the Local Universe. Data were derived with the spectrophotometer PMAS \citep{Roth2005} in the PPAK mode \citep{Verheijen2004,Kelz2006} mounted on the 3.5-meter telescope at the Calar Alto Observatory. The final data release (DR3; \citealp{Sanchez2016}) contains observations of over 600 galaxies spanning all morphological types with redshift $0.005<z<0.03$ in the Local Universe. This release also provides stellar kinematics extracted from the medium-resolution V1200 setup and the $r$-band images from SDSS DR8 \citep{Aihara2011} for an ensemble of 300 galaxies with the stellar mass range from $10^{8.7}$ to $10^{11.9}\,\rm M_{\odot}$. These 300 CALIFA galaxies were modelled using the triaxial Schwarzschild code implemented by \citet{vdB2008}, and the stellar orbit distributions of these CALIFA galaxies were then derived from the models \citep{Zhu2018b}. A bug in this triaxial Schwarzschild code was recently reported \citep{Quenneville2022}, but the stellar orbit distributions based on the original code should not be affected \citep{Thater2022}. Thus we directly used the best-fitting models from \citet{Zhu2018b} in our work.\\

\subsection{Stellar age maps and sample selection}
\label{sec4.2}
\begin{figure}
\begin{centering}
	\includegraphics[width=8.5cm]{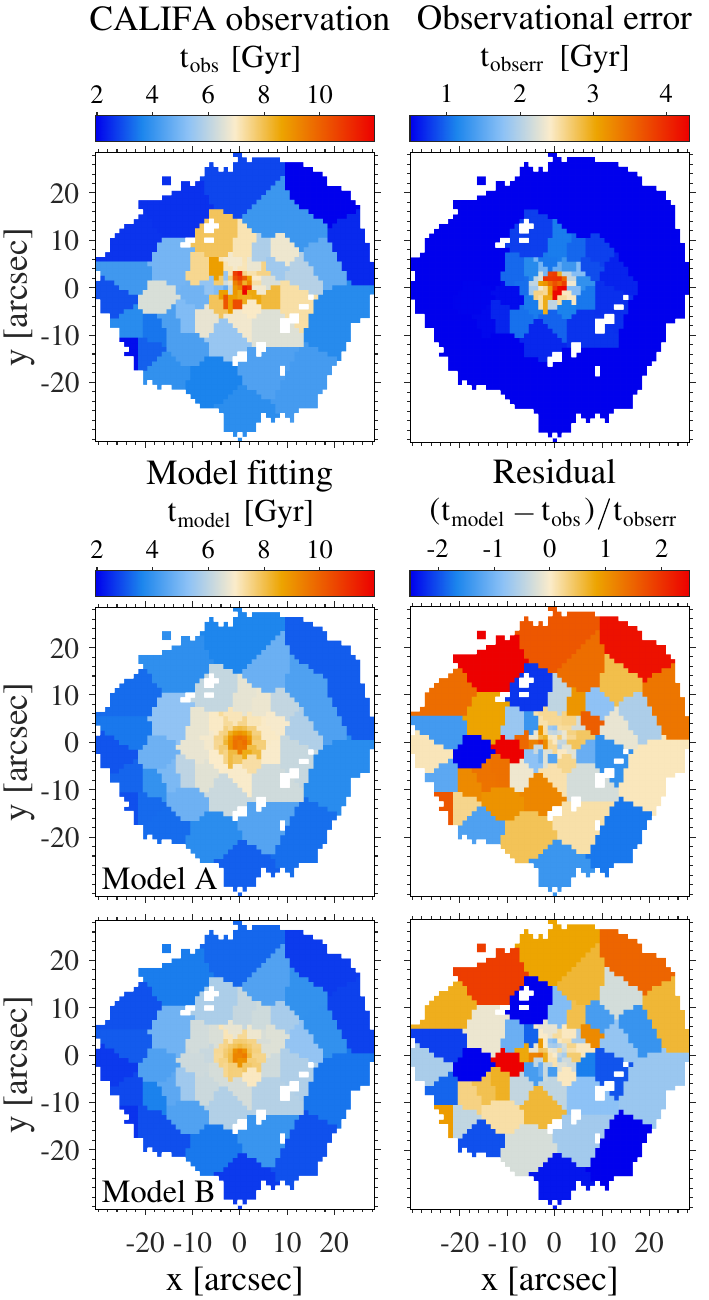}
    \caption{{\bf Binned age map and best-fitting models for NGC 0171}. Top panels: Binned age map $t_{\rm obs}$ and binned error map $t_{\rm obserr}$. Middle panels: Best-fitting age map $t_{\rm model}$ and standardised residuals between mock observations and model fittings $t_{\rm stdres}=(t_{\rm model}-t_{\rm obs})/t_{\rm obserr}$ derived from model A. Bottom: Similar to middle panels, but derived from model B.}
    \label{fig8}
\end{centering}
\end{figure}

The stellar age maps we used originally came from \citet{Zibetti2017}. They derived the stellar population properties of CALIFA galaxies based on a set of five spectral indices and the photometric fluxes in the five SDSS bands, using a Bayesian approach similar to that of \citet{Gallazzi2005}. We matched the catalogues from \citet{Zhu2018b} and \citet{Zibetti2017} and obtained 227 galaxies with both complete standard orbit-superposition models and observational age maps, including 160 spiral galaxies and 67 early-type galaxies. We further excluded spiral galaxies with fewer than 30 apertures in their kinematic maps, which resulted in 113 spiral galaxies with guaranteed data quality and reliable stellar orbit distributions from the orbit-superposition model. We then removed those with obvious asymmetric features in the age maps inspected by eye because our model is point-symmetrised and can never match the asymmetric features. In the end, we obtained a final sample of 82 spiral galaxies with their stellar mass ranging from $10^{8.9}$ to $10^{11.3}\,\rm M_{\odot}$. The surface brightness profiles of our sample within different mass bins are shown in Fig.~\ref{figA1}, and the age profiles of our sample are shown in Fig.~\ref{figA2}. Most of the spiral galaxies in our final sample have negative age gradients. We calculated the luminosity-weighted mean age within and outside half the effective radius $R_{\rm e}/2$ and found that $94\%$ of the galaxies are older in the inner regions and younger in the outer regions. This suggests that the bulges in spiral galaxies, which are spatially concentrated, are mostly older than the disks, which are extended. The 67 early-type galaxies lack obvious age gradients statistically, and we did not tag their orbits with ages. We mention them in the discussion section when we compare with the main results from spiral galaxies by directly taking the average age of the whole galaxy as the age of the bulge. \\

\subsection{Fitting the stellar age maps}
\label{sec4.3}
For each of the 82 spiral galaxies, we took the best-fitting orbit-superposition model and tagged stellar ages to the orbits. Before fitting the age map, we first binned the age map to reduce the high-frequency noise. For each pixel $j$ in a galaxy, an $r$-band luminosity-weighted mean stellar age $t_j$ together with its uncertainty $\sigma(t_j)$ was calculated by \citet{Zibetti2017}. Following the same spatial binning scheme as the kinematic map, the luminosity-weighted binned age could be written as
\begin{equation}
\label{age}
    t_{\rm obs}^i=\frac{\sum_{j=1}^{N_i} L_j t_j}{\sum_{j=1}^{N_i} L_j},\\
\end{equation}
where $L_j$ is the $r$-band luminosity of pixel $j$, and $N_i$ represents the number of pixels in aperture $i$. Thus, the characteristic error of the binned age could be calculated by
\begin{equation}
\label{age-err}
    t_{\rm obserr}^i=\frac{\sqrt{\sum_{j=1}^{N_i} L_j^2 \sigma(t_j)^2}}{\sum_{j=1}^{N_i} L_j}.\\
\end{equation}
The binned age $t_{\rm obs}^i$ obtained in this way should be statistically consistent with that derived from the stacked spectrum of all pixels in each aperture \citep{Ge2019,Ge2021}. However, the surrounding pixels in regions with low S/N were adaptively smoothed in the stellar population analysis by \citet{Zibetti2017}, which results in stellar ages of nearby pixels that are not fully independent. Therefore, the error of the binned age $t_{\rm obserr}^i$ might be underestimated, particularly in the outer part of the galaxy. To address this issue and prevent the dominance of any data point in the $\chi^2$ when fitting the age map, we restricted the minimum value of $t_{\rm obserr}^i$ to be 0.5 Gyr. We illustrate in Appendix~\ref{appendixB} that the age error maps created in this way do not affect our main results. Fig.~\ref{fig8} shows the observational age map $t_{\rm obs}$, the age error map $t_{\rm obserr}$, the best-fitting models $t_{\rm model}$ from models A and B, and the standardised residuals between observations and model fittings for the typical CALIFA galaxy NGC 0171. Both models generally match the observational map, and the standardised residuals are smaller than unity for more than half of the apertures.\\

As described in $\S$~\ref{sec3.3}, we added a restriction of $t_{\lambda_z=0}\ge t_{\lambda_z=1}$ on the age-circularity priors in model B. However, this may not be suitable for galaxies whose bulges are of similar ages or younger than the disks. For the CALIFA sample, we only kept the restriction $t_{\lambda_z=0}\ge t_{\lambda_z=1}$ for galaxies whose inner parts ($r<R_{\rm e}/2$) are 0.5 Gyr older than their outer parts ($r>R_{\rm e}/2$), and we eliminated this restriction for the rest of galaxies. In addition, we followed the same way as introduced in $\S$~\ref{sec3} to construct models A and B for our CALIFA sample. We took the results from model B as our default results.\\

\subsection{Intrinsic stellar age distributions}
\label{sec4.4}
\begin{figure*}
\begin{centering}
	\includegraphics[width=16cm]{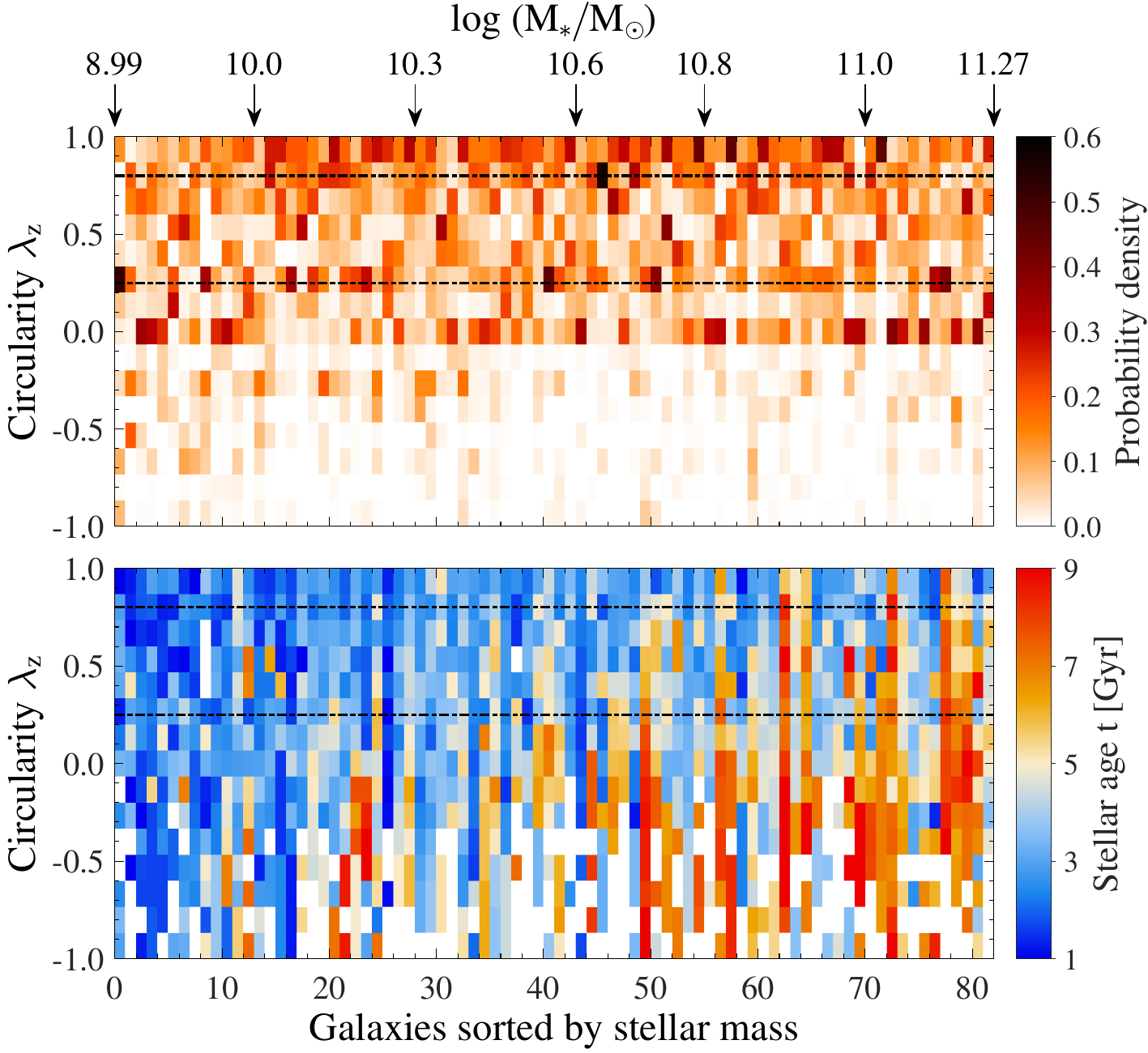}
    \caption{{\bf Probability density distributions of stellar orbits and the distributions of stellar ages within $R_{\rm e}$ vs circularity $\lambda_z$ for each of the 82 CALIFA spirals.} Each column in the top panel shows the orbit probability $P(\lambda_z)$ of a certain circularity bin ($y$-axis) for a galaxy ($x$-axis), as indicated by the colour bar. The sequence of galaxies along the $x$-axis is sorted by increasing stellar mass from left to right, with the positions of some typical stellar mass $\log(M_*/\,\rm M_{\odot})$ shown at the top of the figure. For each galaxy, the average circularity of all stellar orbits is calculated, with the black folded line denoting its variation with stellar mass. The results in the top panel originally come from Figure 2 of \citet{Zhu2018b}, which plots a similar orbit distribution for each of 300 CALIFA galaxies. The bottom panel correspondingly presents the age $t$ of each circularity bin for each galaxy, with bluer pixels indicating younger stellar populations and redder pixels denoting older populations. The horizontal dashed lines in the top and middle panels divide the cold ($\lambda_z\ge0.8$), warm ($0.25<\lambda_z<0.8$), and hot ($\lambda_z\le0.25$, including the counter-rotating orbits) components.}
    \label{fig9}
\end{centering}
\end{figure*}

Now we had the stellar orbit distributions with the stellar ages tagged to the orbits for all the 82 spiral galaxies. For each galaxy, we took all stellar orbits within the effective radius $R_{\rm e}$ and divided them into 15 bins of circularity. Then we calculated the probability density of orbits and luminosity-weighted mean stellar ages in each bin. We thus obtained the probability density distributions of orbits $P(\lambda_z)$ and the intrinsic distributions of stellar ages $t(\lambda_z)$ for each galaxy. We also calculated the average circularity of all stellar orbits for each galaxy. The probability density of the orbits is the same as presented in \citet{Zhu2018b}, and the total orbit weight within $R_e$ for each galaxy was normalised to unity.\\

We present $P(\lambda_z)$ and $t(\lambda_z)$ for all the 82 CALIFA spiral galaxies in Fig.~\ref{fig9}, with the galaxies sorted by increasing stellar mass from left to right. The top panel shows the stellar orbit distributions $P(\lambda_z)$, and the bottom panel shows the intrinsic stellar age distributions $t(\lambda_z)$ derived from model B. The stars on all orbits become systematically older with increasing stellar mass, with a more significant increase for stars on random-motion dominated orbits than those on rotation-dominated orbits. For low-mass galaxies, stars on all orbits are similarly young. For high-mass galaxies, stars on random-motion dominated orbits become significantly older in most cases, while the ages of stars on rotation-dominated orbits span a wide range, which means that disks are much older in some cases and are young in other cases.\\

\subsection{Stellar ages of different dynamical components}
\label{sec4.5}
\begin{figure}
\begin{centering}
    \includegraphics[width=8.5cm]{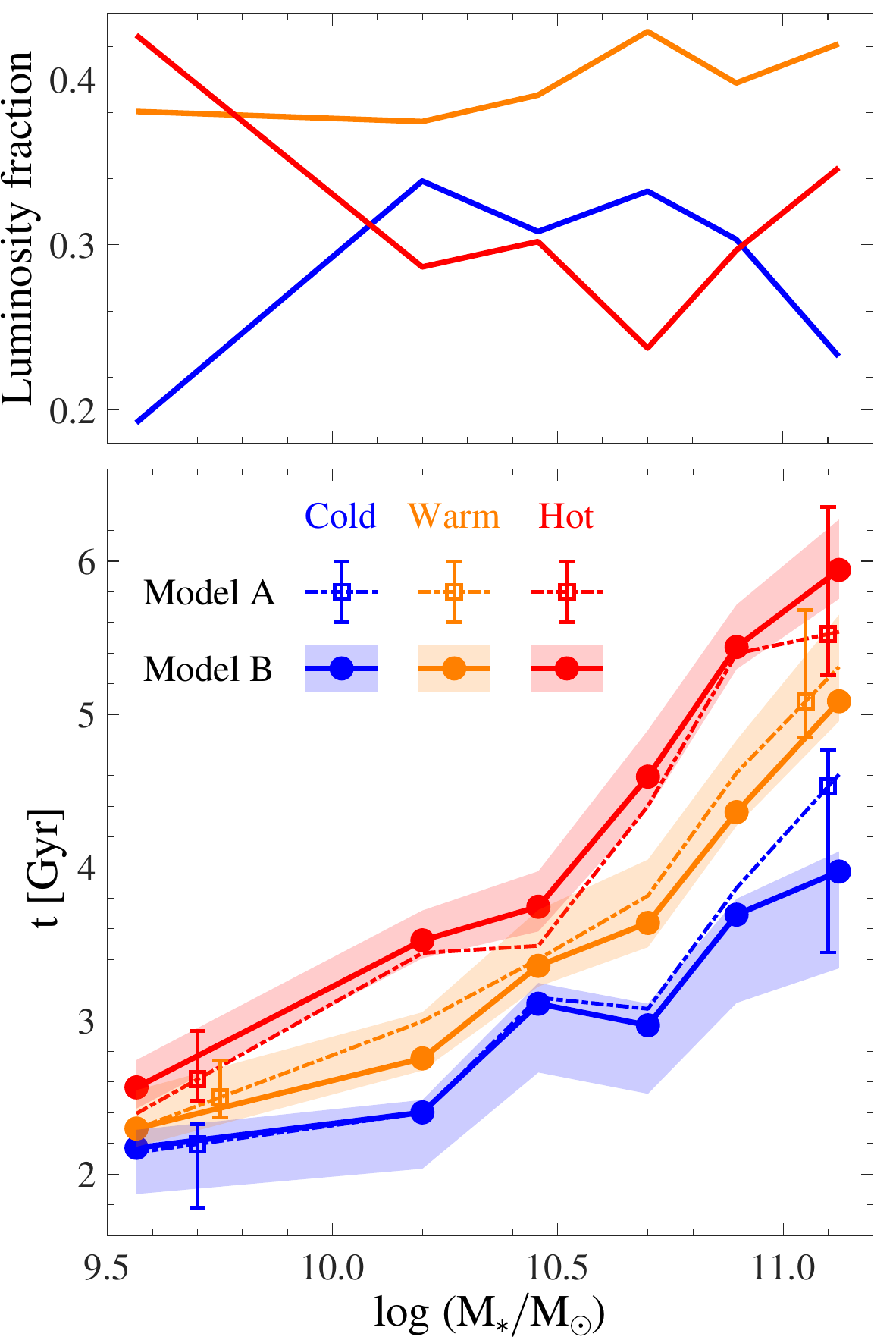}
    \caption{{\bf Luminosity-weighted orbit fractions and mean stellar ages of the cold, warm, and hot components within $R_{\rm e}$ as a function of the galaxy stellar mass.} In the top panel, the blue, orange, and red lines represent the luminosity-weighted fractions of the cold, warm, and hot components in different mass bins. In the bottom panel, the coloured solid lines with dots denote the mean ages of the cold, warm, and hot components within different mass bins derived from model B. The coloured shadows represent the overall uncertainties of model B. For comparison, the mean ages of the cold, warm, and hot components derived from model A are shown by the coloured dashed lines, with the coloured error bars representing the typical overall uncertainties of model A at the low-mass and high-mass ends.}
    \label{fig10}
\end{centering}
\end{figure}

\begin{figure}
\begin{centering}
    \includegraphics[width=8.5cm]{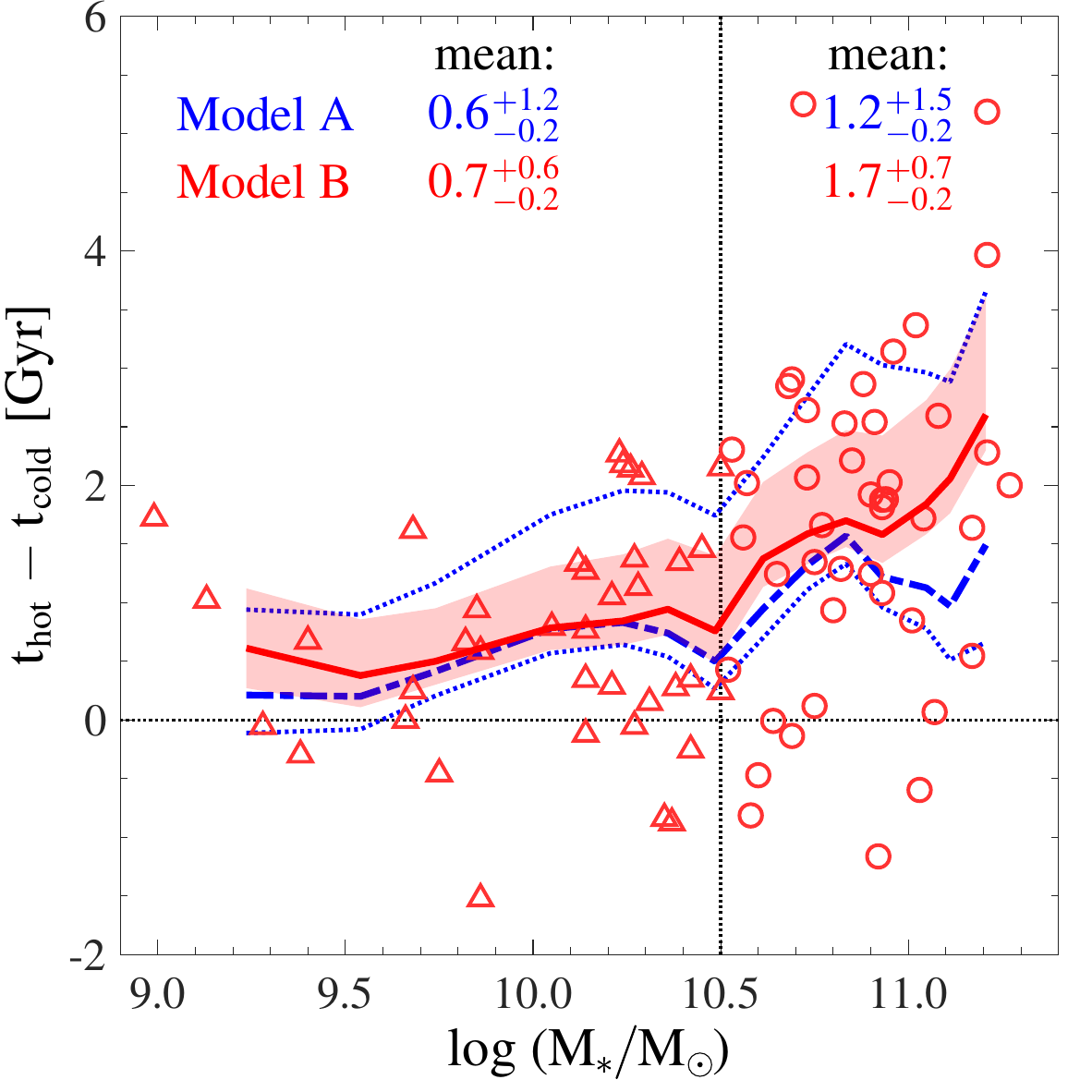}
    \caption{{\bf Age difference between the hot and cold orbit components $t_{\rm hot}-t_{\rm cold}$ within $R_{\rm e}$ vs stellar mass.} The red markers indicate $t_{\rm hot}-t_{\rm cold}$ derived from model B for each galaxy in our sample, with the triangles representing low-mass spirals ($M_*\le10^{10.5}\,\rm M_{\odot}$) and the circles representing high-mass spirals ($M_*>10^{10.5}\,\rm M_{\odot}$). The corresponding red curve represents the moving average of $t_{\rm hot}-t_{\rm cold}$. The red shadow denotes the overall uncertainties of model B. For comparison, the moving average of $t_{\rm hot}-t_{\rm cold}$ and the overall uncertainties derived from model A are shown by the dashed and dotted blue lines. The horizontal dotted line means $t_{\rm hot}=t_{\rm cold}$, while the vertical dotted line denotes $M_*=10^{10.5}\,\rm M_{\odot}$. The mean values of $t_{\rm hot}-t_{\rm cold}$ for low-mass spirals and high-mass spirals are shown by the texts.}
    \label{fig11}
\end{centering}
\end{figure}

For each galaxy, we divided the orbits within $R_{\rm e}$ into three components: cold ($\lambda_z\ge0.8$), warm ($0.25<\lambda_z<0.8$), and hot ($\lambda_z\le0.25$), as indicated in Fig.~\ref{fig9}. We then calculated the luminosity-weighted orbit fraction and mean stellar age of each component. According to the morphological analysis of different orbit components \citep{Zhu2018a}, these galaxies do not have obvious counter-rotating disks, and the counter-rotating orbits usually exist in the inner parts of galaxies. Both hot and counter-rotating orbits contribute to the bulges in these galaxies, and the cold orbits make up the majority of the disks, while the warm orbits contribute either to bulges or disks. We thus took the cold component as the dynamical disk and the hot component, which includes both hot and counter-rotating orbits, as the dynamical bulge in this paper.\\

We divided our sample of 82 galaxies into six mass bins, with intervals defined by $\log(M_*/\,\rm M_{\odot})=8.9$, $10.0$, $10.3$, $10.6$, $10.8$, $11.0$, and $11.3$. For each mass bin, we calculated the average fractions of the cold, warm, and hot components, and we present them in the top panel of Fig.~\ref{fig10}. The cold-orbit fraction peaks in galaxies with intermediate mass ($M_*=10^{10}\sim10^{11}\,\rm M_{\odot}$), and decreases at both the low-mass ($M_*<10^{10}\,\rm M_{\odot}$) and high-mass ($M_*>10^{11}\,\rm M_{\odot}$) ends, while the hot-orbit fraction correspondingly increases at the low- and high-mass ends. In the bottom panel of Fig.~\ref{fig10}, we present the average stellar ages of the cold, warm, and hot components ($\overline{t_{\rm cold}}$, $\overline{t_{\rm warm}}$ , and $\overline{t_{\rm hot}}$) in each bin. The overall uncertainties are also shown in the panel, which includes statistical uncertainties and systematic biases. For a bin with $N$ galaxies, the statistical uncertainty contributes $\sigma_{\rm stat}/\sqrt{N}$ to the upper and lower overall uncertainties, while the systematic bias $\overline{\mathcal{D}}$ is added to the upper or lower overall uncertainties, depending on its sign. For example, if $\overline{\mathcal{D}}<0$, the model underestimates the value, so that the upper overall uncertainty is increased by $|\overline{\mathcal{D}}|$ to recover the true value. According to our tests in $\S$~\ref{sec3.4}, the statistical uncertainties of each galaxy for the ages of the cold, warm, and hot components are $\sigma_{\rm stat}=11\%$, $3\%$, and $8\%$, respectively, and the corresponding systematic biases are $\overline{\mathcal{D}}=14\%$, $-9\%$, and $-3\%$. For the 66 galaxies whose inner parts of the age maps are 0.5 Gyr older than the outer parts (see $\S$~\ref{sec4.3}), we treated them as similar to the mock galaxies and used the test results to estimate their overall uncertainties. For the other 14 galaxies that are not similar to the mock galaxies, we set a secure statistical uncertainty of $30\%$ for the stellar ages of each component.\\

From Fig.~\ref{fig10}, we find that most galaxies have younger stars in dynamically colder components, the stars in the cold disks are the youngest, and those in the hot bulges are the oldest ($t_{\rm cold}<t_{\rm warm}<t_{\rm hot}$). All three components become older with increasing stellar mass, which is consistent with the scenario that more massive spirals tend to form both their bulges and disks earlier. Furthermore, the age of the hot component $t_{\rm hot}$ obviously increases faster than that of the cold component $t_{\rm cold}$. At $M_*<10^{10}\,\rm M_{\odot}$, the mean ages of the cold, warm, and hot components derived from model B ($\overline{t_{\rm cold}}$, $\overline{t_{\rm warm}}$ , and $\overline{t_{\rm hot}}$) are $\sim2.2$ Gyr, $\sim2.3$ Gyr, and $\sim2.6$ Gyr, respectively, while at $M_*>10^{11}\,\rm M_{\odot}$, the values increase to $\sim4.0$ Gyr, $\sim5.1$ Gyr, and $\sim5.9$ Gyr. For comparison, we also present the results from model A. Model A provides similar mean ages as model B, but with larger uncertainties.\\

In order to further investigate the age difference between dynamically hot bulges and dynamically cold disks, we plot $t_{\rm hot}-t_{\rm cold}$ versus stellar mass $\log(M_*/\,\rm M_{\odot})$ in Fig.~\ref{fig11}. We calculated the moving average of $t_{\rm hot}-t_{\rm cold}$, and the uncertainties of the moving average have a similar definition as those in Fig.~\ref{fig10}. In model B, the age difference between the hot and cold components increases with stellar mass from $t_{\rm hot}-t_{\rm cold}\sim 0.5$ Gyr at the low-mass end to $t_{\rm hot}-t_{\rm cold}\sim2$ Gyr at the high-mass end. This increase of $t_{\rm hot}-t_{\rm cold}$ becomes rapid around a critical mass of $M_*=10^{10.5}\,\rm M_{\odot}$. Thus, we separated our sample into low-mass ($M_*\le10^{10.5}\,\rm M_{\odot}$) and high-mass ($M_*>10^{10.5}\,\rm M_{\odot}$) spiral galaxies. For low-mass spirals, the age difference between the cold and hot components $t_{\rm hot}-t_{\rm cold}$ is usually smaller than 2 Gyr, with a mean value of $\overline{t_{\rm hot}-t_{\rm cold}}=0.7_{-0.2}^{+0.6}$ Gyr; while for high-mass spirals, the age difference spans a wide range, with a mean value of $\overline{t_{\rm hot}-t_{\rm cold}}=1.7_{-0.2}^{+0.7}$ Gyr. The scatter of the age difference in low-mass galaxies is relatively small, with $\sigma(t_{\rm hot}-t_{\rm cold})=0.9$ Gyr, while it increases to 1.4 Gyr in high-mass galaxies.\\

For comparison, we also present the results from model A in Fig.~\ref{fig11}. Model A also shows a similar trend that $t_{\rm hot}-t_{\rm cold}$ increases with stellar mass, but with larger uncertainties. About $80\%$ of galaxies (66 out of 82) in model B and $79\%$ of galaxies (65 out of 82) in model A have bulges that are older than disks ($t_{\rm hot}>t_{\rm cold}$). Although we set strong age-circularity priors in model B, the fraction of galaxies whose bulges are older than disks is not significantly different from model A, which started from a uniform prior (see $\S$~\ref{sec3.3} and $\S$~\ref{sec4.3}).\\

Combing Fig.~\ref{fig9}, Fig.~\ref{fig10}, and Fig.~\ref{fig11}, we infer the assembly histories of CALIFA spiral galaxies. Most spirals ($80\%$) have bulges that are older on average than their disks. Low-mass spirals have relatively young bulges and disks, with bulges being slightly older. This is consistent with the scenario found in the cosmological simulation TNG50: less massive spirals, especially those with $M_*\lesssim 10^{10}\,\rm M_{\odot}$, form their bulges and disks simultaneously with continuous star formation from high to intermediate redshift, with the star formation in disks lasting longer to the present day, resulting in slightly younger disks on average (Zhang et al. in prep). High-mass spirals have old bulges, and their disks are generally younger than their bulges, but with a wide range of stellar ages. The bulges in massive galaxies ($M_*\gtrsim 10^{10.5}\,\rm M_{\odot}$) are likely formed in the star formation burst at high redshift (Zhang et al. in prep), while the disks are built after the bulge formation and could be destroyed by mergers but might be rebuilt afterwards (e.g. \citealp{Springel2005,Robertson2006,Hopkins2009}). Thus the wide range of disk stellar ages could be caused by their different merger histories. We also note that in some galaxies, the bulges are younger than disks, which could be caused by a few processes such as environmental gas stripping in the outer regions (e.g. \citealp{Schaefer2019,Owers2019,Lin2019,Ding2023}) and stellar winds (e.g. \citealp{Pipino2008,Pipino2010,Zibetti2020}).\\

\section{Discussion}
\label{sec5}
\begin{figure*}
\begin{centering}
    \includegraphics[width=12cm]{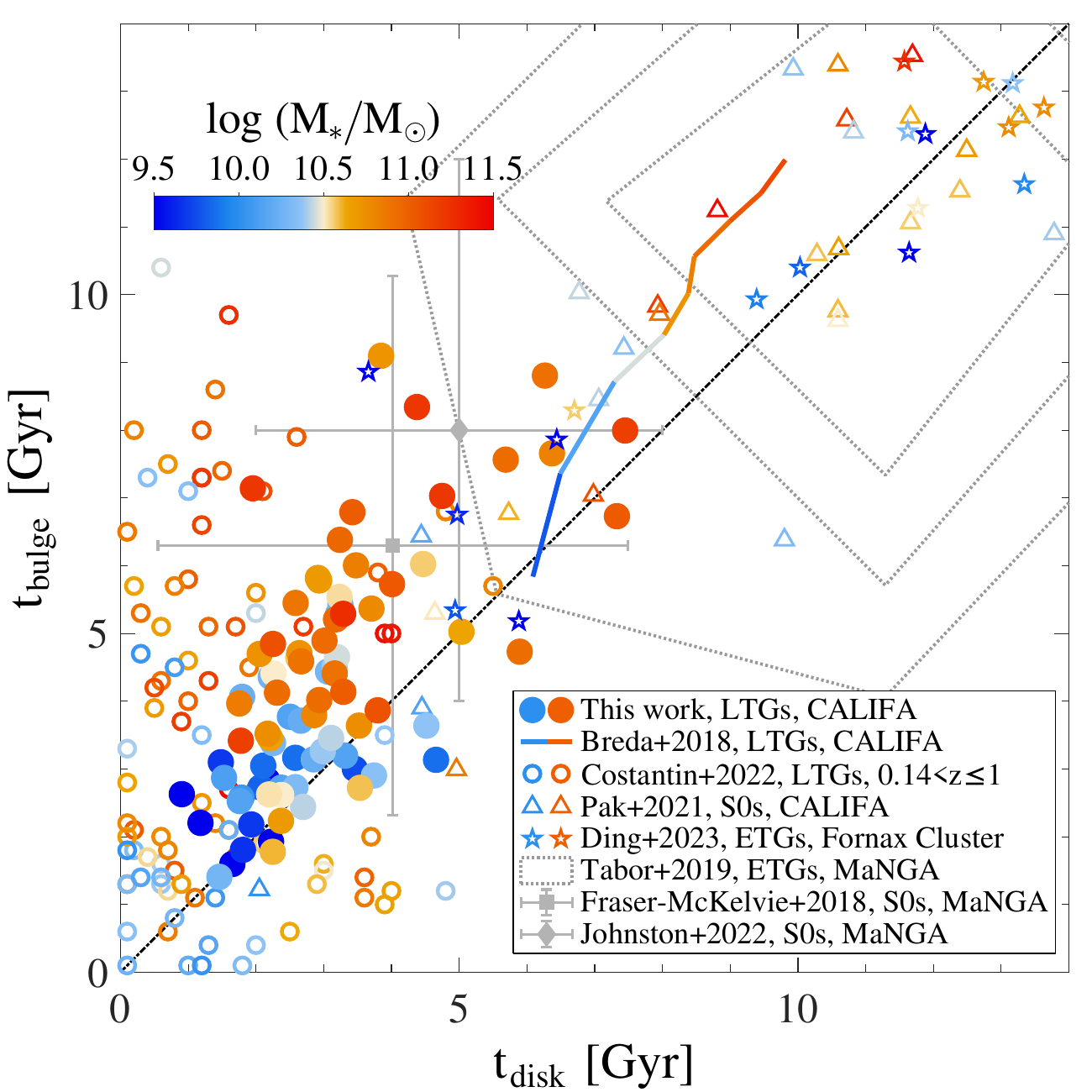}
    \caption{{\bf One-to-one comparison of bulge ages $t_{\rm bulge}$ and disk ages $t_{\rm disk}$ from different works.} The dots represent 82 CALIFA spirals in our sample, with the bulge and disk ages corresponding to the ages of the hot and cold components within $R_{\rm e}$ derived from model B ($t_{\rm bulge}$=$t_{\rm hot}$, $t_{\rm disk}$=$t_{\rm cold}$). The coloured folded line illustrates the variation of mean bulge ages with mean disk ages for 135 CALIFA spiral galaxies from \citet{Breda2018}. The circles indicate 91 spiral galaxies at redshift $0.14<z\le1$ from \citet{Costantin2022}. The triangles denote 29 CALIFA S0 galaxies (shifted to 2 Gyr younger for both bulges and disks) from \citet{Pak2021}. The pentacles denote 18 early-type galaxies in the Fornax Cluster from \citet{Ding2023}. The stellar mass of galaxies are indicated by the colour bar. The region covered by the grey dashed lines (shifted to 4 Gyr younger for both bulges and disks) roughly shows the density distributions of 272 MaNGA early-type galaxies from \citet{Tabor2019}. The grey squares and diamonds represent the mean bulge ages and disk ages in the samples of 279 MaNGA S0 galaxies \citep{FraserMcKelvie2018} and 78 MaNGA S0 galaxies \citep{Johnston2022b}, respectively, with the corresponding error bars representing the standard deviations of bulge ages and disk ages.}
    \label{fig12}
\end{centering}
\end{figure*}
\begin{figure*}
\begin{centering}
    \includegraphics[width=12cm]{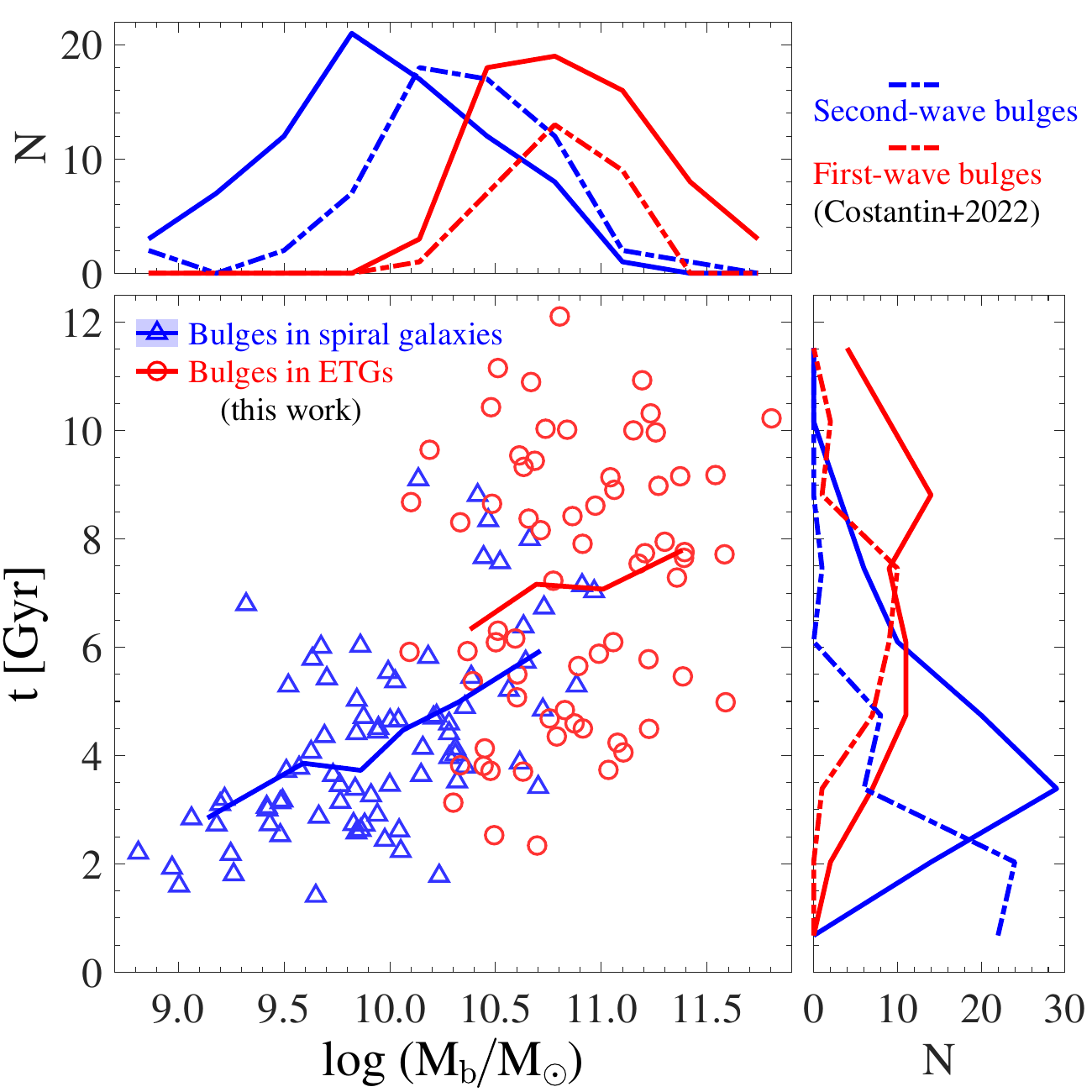}
    \caption{{\bf Bulge ages vs bulge mass for spiral galaxies and early-type galaxies.} In the main panel, the blue triangles represent 82 bulges in spiral galaxies in our sample, and their ages correspond to the ages of the hot component within $R_{\rm e}$ derived from model B. The red circles denote 67 bulges in early-type galaxies in our contrast sample, and their ages correspond to the luminosity-weighted mean ages calculated from the age maps. The solid blue line with the shadow represents the mean bulge ages and uncertainties of spiral galaxies within different mass bins, and the solid red line denotes the mean bulge ages of early-type galaxies within different mass bins. In the top and right panels, the corresponding coloured solid lines represent the marginalised distributions of bulge mass and bulge ages in our work. The dashed lines denote the distributions from \citet{Costantin2022}, and the blue lines represent the second-wave bulges, and the red lines represent the first-wave bulges.}
    \label{fig13}
\end{centering}
\end{figure*}

\subsection{Bulge versus disk stellar ages}
\label{sec5.1}
Several studies in the literature have aimed to obtain the stellar ages of bulges and disks separately (e.g. \citealp{Breda2018,FraserMcKelvie2018,Tabor2019,Pak2021,Johnston2022b,Costantin2022,Ding2023}), and we present them in this section. The bulge ages and disk ages of 135 CALIFA spiral galaxies were derived by \citet{Breda2018}, who used the surface photometry, the spectral modelling of IFS data based on the spectral fitting software STARLIGHT \citep{CidFernandes2005}, and the post-processing tool REMOVEYOUNG \citep{Gomes2016}. The stellar ages of bulges and disks were derived for 91 spiral galaxies with $M_*>10^{10}\,\rm M_{\odot}$ at redshift $0.14<z\le1$ \citep{Costantin2022} from the spectral energy distributions (SED) with spectral resolution $R\sim50$ up to redshift $z=1$ \citep{Costantin2021}, taking full advantage of the spectral information provided by SHARDS \citep{PerezGonzalez2013} and the spatial resolution given by HST/CANDELS \citep{Koekemoer2011,Grogin2011}. By preforming the spectroscopic decomposition method presented in \citet{Tabor2017}, the stellar ages of bulges and disks were derived for 272 MaNGA early-type galaxies \citep{Tabor2019} and 34 CALIFA S0 galaxies \citep{Pak2021}. Based directly on photometric decomposition, 279 MaNGA S0 galaxies were decomposed, and then their bulge ages and disk ages were derived \citep{FraserMcKelvie2018}. The bulge ages and disk ages of 78 MaNGA S0 galaxies were derived \citep{Johnston2022b} via the 2D multi-band image fitting software BUDDI \citep{Johnston2017}. Based on the same population-orbit superposition method, the stellar ages of dynamically cold disks and hot bulges were derived for 18 early-type galaxies in the Fornax Cluster by \citet{Ding2023}, with the bulge and disk defined in a similar way as in our work.\\

In Fig.~\ref{fig12} we plot the bulge ages $t_{\rm bulge}$ versus the disk ages $t_{\rm disk}$ from the above works and compare them with our results. For low-mass spiral galaxies ($M_*\le10^{10.5}\,\rm M_{\odot}$) in CALIFA, both bulges and disks are relatively young, with bulges being $\sim 1$ Gyr older; while for high-mass spiral galaxies ($M_*>10^{10.5}\,\rm M_{\odot}$) in CALIFA, both bulges and disks become older, with bulges becoming $\sim 2$ Gyr older than disks. The mass dependence of $t_{\rm bulge}-t_{\rm disk}$ in our work is exactly the same as in \citet{Breda2018}, while their bulge ages and disk ages systematically exceed ours, which might be caused by the removal of young stellar populations in their work. The results of spiral galaxies are consistent with the works on S0 galaxies, where the bulges are slightly older (\citealp{Pak2021}, CALIFA) or 2-3 Gyr older on average (\citealp{FraserMcKelvie2018, Johnston2022b}, MaNGA) than disks. However, early-type galaxies in MaNGA have similar bulge ages and disk ages, as suggested by \citet{Tabor2019}. The comparison might be biased because the definitions of bulges and disks are not exactly the same in the different methods.\\

The stellar ages of bulges and disks in \citet{Ding2023} were obtained using the same method as we used here, but for early-type galaxies in the Fornax cluster. The age difference between bulges and disks was found to be related with the galaxy infall time into the cluster. The ancient infallers tend to have similar bulge ages and disk ages as a result of gas removal in the cluster environment, while the intermediate and recent infallers have $\sim 2$ Gyr older bulges than disks, similar to the high-mass spirals in CALIFA.\\

For galaxies with slightly higher redshift, the majority of spiral galaxies ($85\%$) also have bulges older than disks, as shown in \citet{Costantin2022}, and the age difference $t_{\rm bulge}-t_{\rm disk}$ increases with stellar mass, which is consistent with our results. However, both $t_{\rm disk}$ and $t_{\rm bulge}$ in their work are smaller than ours, which may have two reasons. Firstly, their galaxies are at redshift $0.14<z\le1$ and our CALIFA galaxies are at $z\sim0$, which will cause both their bulges and disks to be younger than ours. Secondly, our disks are defined within $R_{\rm e}$ of the galaxies, while their disks might cover larger radii. Because we expect stars in the outer disk to be younger than those in the inner disk with inside-out growth (e.g. \citealp{GonzalezDelgado2015,Goddard2017,Ma2017}), the average disk age will be different with different spatial coverage. When we increase the spatial coverage of the disks from $R_{\rm e}$ to 1.5 $R_{\rm e}$, the average disk age we obtain is $\sim 0.2$ Gyr younger.\\

\subsection{Bulge stellar age versus mass}
\label{sec5.2}
We further investigated the dependence of the bulge age $t_{\rm bulge}$ on the bulge mass $\log(M_{\rm b}/\,\rm M_{\odot})$. We included 82 CALIFA spiral galaxies and 67 CALIFA early-type galaxies in our sample. For all CALIFA galaxies, the stellar orbit distributions were derived from the orbit-superposition method \citep{Zhu2018b}, and we defined the bulges in early-type galaxies in the same way as in spiral galaxies. The early-type galaxies show no obvious stellar age gradients, which indicates that the stellar ages of bulges and disks are statistically similar. We thus take the luminosity-weighted mean ages calculated from the age maps to represent the bulge ages of these early-type galaxies. As shown in Fig.~\ref{fig13}, the stellar ages of bulges increase with bulge mass for both spiral galaxies and early-type galaxies. At similar bulge mass ($M_{\rm b}\sim10^{10.5}\,\rm M_{\odot}$), bulges in early-type galaxies span a wide range of ages and are similarly old as those in spiral galaxies.\\

In the top and right panels of Fig.~\ref{fig13}, we compare the marginalised distributions of the bulge mass and the bulge ages with those from \citet{Costantin2022}. They found a bimodal distribution of the bulge stellar age in their sample, and defined the bulges formed at redshift $z>3$ ($t>11.7$ Gyr) as the first-wave bulges and the bulges formed after $z<3$ as the second-wave bulges. They claimed that the first-wave bulges built up fast in the early Universe through dissipative processes, probably evolving to the well-known $z\sim2$ ($t\sim10.5$ Gyr) red nuggets, while the second-wave bulges form at $z=1$-2 ($t=7.9$-10.5 Gyr), and both types of bulges tend to acquire their disks by $z\sim0.5$ ($t\sim5.2$ Gyr). The mass and age distributions in the bulges of our spiral galaxies are similar to the second-wave bulges, except that there are more low-mass bulges in our sample, which is likely caused by the selection bias, and the bulge ages of our spiral galaxies are older than the second-wave bulges, which is likely due to the reason we explained when introducing Fig.~\ref{fig12}. However, we lack spiral galaxies with bulges corresponding to the first-wave bulges. Instead, distributions of the the bulges in our early-type galaxies are similar to those of the first-wave bulges in both bulge mass and ages. This indicates that the bulges in local early-type galaxies may have the same origin as the first-wave bulges in massive spiral galaxies at higher redshift. Firstly, a spiral galaxy containing a first-wave bulge at high redshift may quench later and become a present-day early-type galaxy. Secondly, if a first-wave bulge formed in the early Universe does not acquire a disk, it could also become a present-day early-type galaxy.\\

When spiral and early-type galaxies are combined, the population of bulges in CALIFA galaxies is generally consistent with the bimodality distribution of bulges in spiral galaxies at higher redshift \citep{Costantin2022}.\\

\section{Summary}
We adopted an innovative technique combining stellar kinematics and stellar populations, the population-orbit superposition method, to study the stellar ages of the cold, warm, and hot components for CALIFA spiral galaxies. We first evaluated the capabilities of this method on 36 mock data sets with CALIFA-like resolution created from Auriga simulated galaxies. By carefully considering the initial priors of Bayesian analysis in fitting the age maps (model B), the recovered mean ages of the cold ($\lambda_z\ge0.8$) and hot ($\lambda_z\le0.25$) components always match the true values within 0.7 Gyr, with observational errors ranging from $5\%$ to $20\%$. However, for Bayesian analysis starting from uniform priors (model A), the recovered mean ages of the cold and hot components could be biased by more than 1 Gyr, with observational errors of $15\%$ and $20\%$.\\

We applied models A and B to 82 CALIFA spiral galaxies and obtained results consistent with each other. We took model B as the default, and our main findings are as follows.\\

\begin{enumerate}
\item Most spirals ($80\%$) in our sample have $t_{\rm hot}>t_{\rm cold}$, which indicates that these galaxies form their bulges first and acquire their disks later, and/or their disks experience more prolonged and extensive period of star formation.

\item The stellar ages of the cold, warm, and hot components ($t_{\rm hot}$, $t_{\rm warm}$ and $t_{\rm cold}$) all increase with galaxy stellar mass. The mean ages of the cold, warm, and hot components ($\overline{t_{\rm hot}}$, $\overline{t_{\rm warm}}$ , and $\overline{t_{\rm cold}}$) increase from $\overline{t_{\rm cold}}\sim2.2$ Gyr, $\overline{t_{\rm warm}}\sim2.3$ Gyr, and $\overline{t_{\rm hot}}\sim2.6$ Gyr for galaxies with $M_*<10^{10}\,\rm M_{\odot}$ to $\overline{t_{\rm cold}}\sim4.0$ Gyr, $\overline{t_{\rm warm}}\sim5.1$ Gyr, and $\overline{t_{\rm hot}}\sim5.9$ Gyr for galaxies with $M_*>10^{11}\,\rm M_{\odot}$.

\item The difference in stellar ages between cold and hot components $t_{\rm hot}-t_{\rm cold}$ increases with the galaxy stellar mass. The hot component is $0.7_{-0.2}^{+0.6}$ Gyr older on average than the cold component for low-mass spirals ($10^{8.9}\,\rm M_{\odot}<M_*\le10^{10.5}\,\rm M_{\odot}$), while it becomes $1.7_{-0.2}^{+0.7}$ Gyr older on average than the cold component for high-mass spirals ($10^{10.5}\,\rm M_{\odot}<M_*<10^{11.3}\,\rm M_{\odot}$).
\end{enumerate}

The systematic change in stellar ages in bulges and disks from low- to high-mass galaxies indicates their different formation histories, which is generally consistent with the scenario found in the cosmological simulation TNG50. Low-mass spirals, especially those with $M_*\lesssim 10^{10}\,\rm M_{\odot}$, form their bulges and disks simultaneously with continuous star formation from high to intermediate redshift, with the star formations on disks lasting longer to the present day, resulting in slightly younger disks on average. The bulges in massive galaxies ($M_*\gtrsim 10^{10.5}\,\rm M_{\odot}$) are likely formed in the star formation burst at high redshift, while the disks are built after the bulge formation and could be destroyed by mergers, but are rebuilt afterwards. Various dry/wet merger histories could cause the wide range of disk ages we found in massive spiral galaxies.\\

\section*{Acknowledgements}
The authors acknowledge the Tsinghua Astrophysics High-Performance Computing platform at Tsinghua University for providing computational and data storage resources. The research presented here is partially supported by the National Natural Science Foundation of China under grant No. Y945271001, and CAS Project for Young Scientists in Basic Research under grant No. YSBR-062. LC acknowledges financial support from Comunidad de Madrid under Atracci\'on de Talento grant No. 2018-T2/TIC-11612 and from grant No. PGC2018-093499-B-I00 by the Spanish Ministry of Science and Innovation/State Agency of Research MCIN/AEI/ 10.13039/501100011033. GvdV acknowledges funding from the European Research Council (ERC) under the European Union's Horizon 2020 research and innovation programme under grant agreement No. 724857 (Consolidator Grant ArcheoDyn).\\

\begin{appendix}

\section{The surface brightness profiles and age profiles for CALIFA spiral galaxies}
\label{appendixA}
\begin{figure*}
\begin{centering}
	\includegraphics[width=16cm]{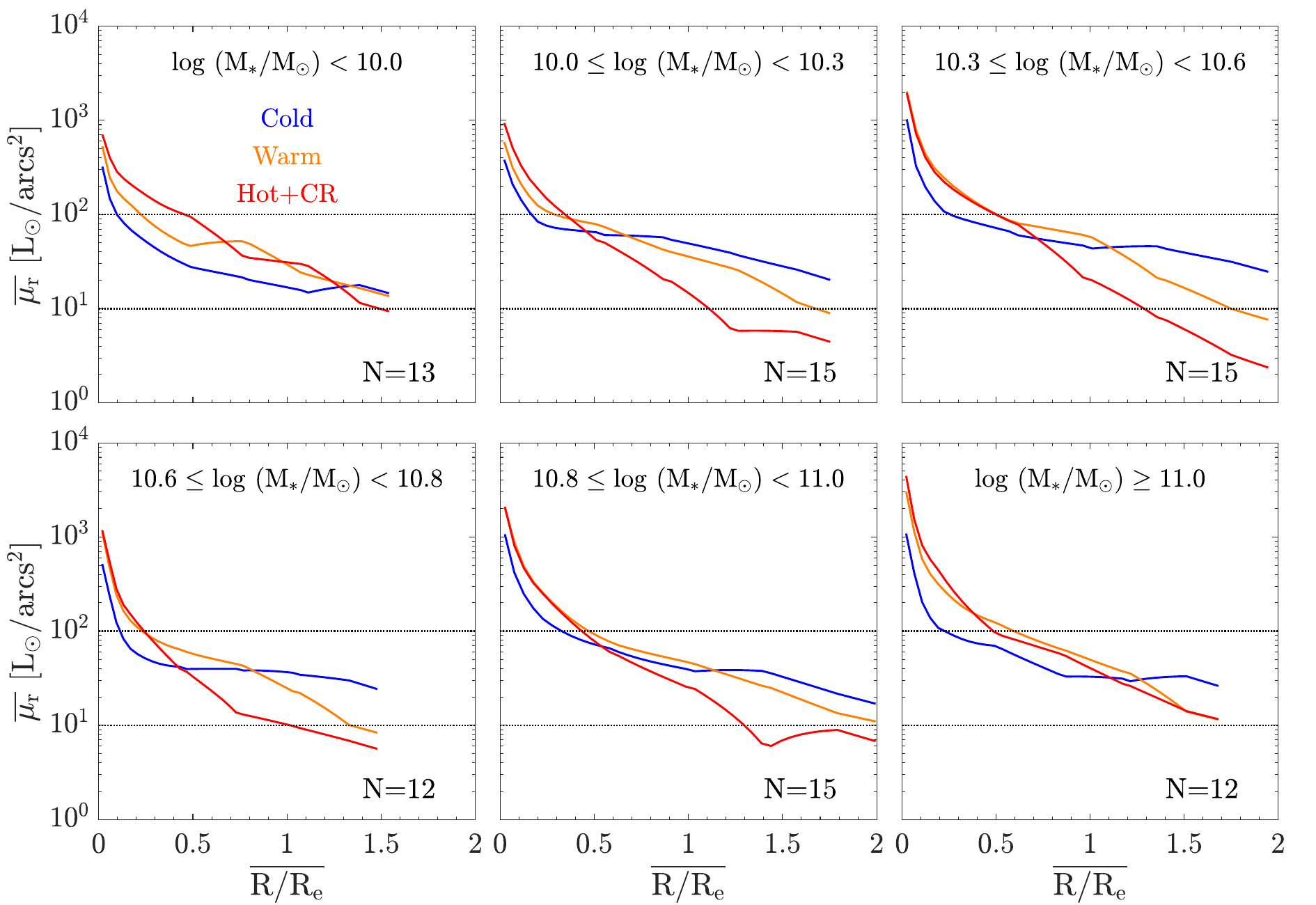}
    \caption{{\bf Binned r-band surface brightness profiles $\overline{\mu_{\rm r}}$ vs radius $\overline{R/R_{\rm e}}$.} Our sample is divided into six mass bins with intervals defined by $\log(M_*/\,\rm M_{\odot})=8.9$, $10.0$, $10.3$, $10.6$, $10.8$, $11.0$, and $11.3$. In each panel, we bin the profiles of galaxies within a certain mass bin for cold ($\lambda_z\ge0.8$, blue lines), warm ($0.25<\lambda_z<0.8$, orange lines), and hot ($\lambda_z\le0.25$, red lines) components separately. The number of galaxies $N$ in each bin is shown by the text. The horizontal dotted lines in each panel represent the values of $10^1$ and $10^2$.}
    \label{figA1}
\end{centering}
\end{figure*}

\begin{figure*}
\begin{centering}
	\includegraphics[width=16cm]{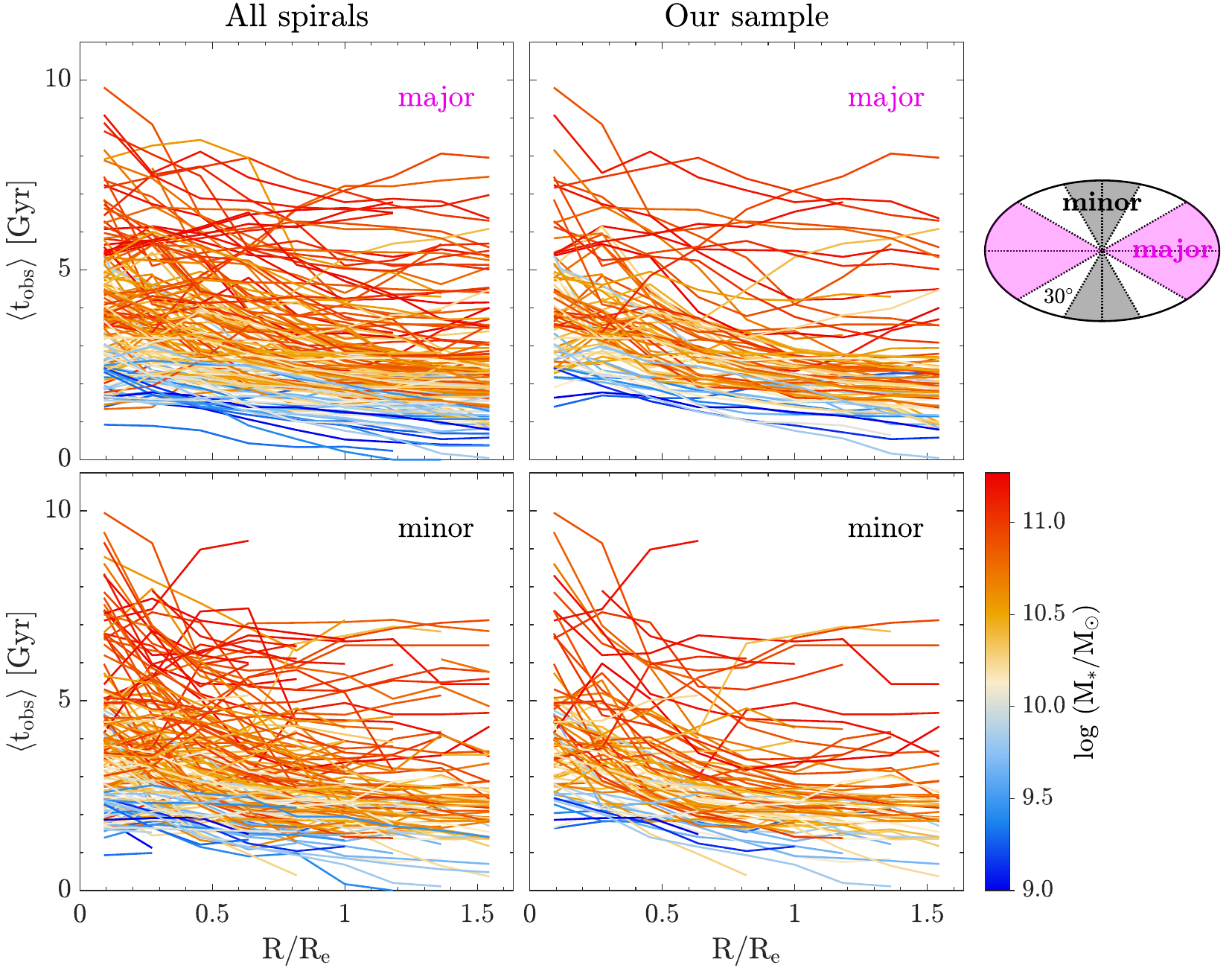}
    \caption{{\bf Luminosity-weighted age profiles $\langle t_{\rm obs}\rangle$ along the major and minor axes vs radius $R/R_{\rm e}$.} The left panels show the profiles of all 160 spiral galaxies by matching the catalogues from \citet{Zhu2018b} and \citet{Zibetti2017}, while the right panels show the profiles of 82 spirals in our final sample. The ellipse on the right of the figure illustrates how we calculate the age profile: for each galaxy, we find its major axis from the image and divide the image into different regions (separated by the dotted lines) by a step of $30^{\circ}$ starting from the major axis. Then we calculate the luminosity-weighted age profile within $\pm30^{\circ}$ regions away from the major axis (magenta shadow, corresponding to the top panels) and the minor axis (grey shadow, corresponding to the bottom panels). Each fold line in the panels indicates a galaxy, with its stellar mass $\log(M_*/\,\rm M_{\odot})$ indicated by the colour bar.}
    \label{figA2}
\end{centering}
\end{figure*}

In Fig.~\ref{figA1} we plot the r-band surface brightness profiles $\overline{\mu_{\rm r}}$ versus radius $\overline{R/R_{\rm e}}$ for the cold, warm, and hot components within different mass bins, based on the results from \citet{Zhu2018b}. In Fig.~\ref{figA2} we present the luminosity-weighted age profiles $\langle t_{\rm obs}\rangle$ along the major and minor axes versus radius $R/R_{\rm e}$, which are directly calculated based on the age maps from \citet{Zibetti2017}.\\

\section{A new version of age error maps and the corresponding results}
\label{appendixB}
\begin{figure}
\begin{centering}
    \includegraphics[width=8.5cm]{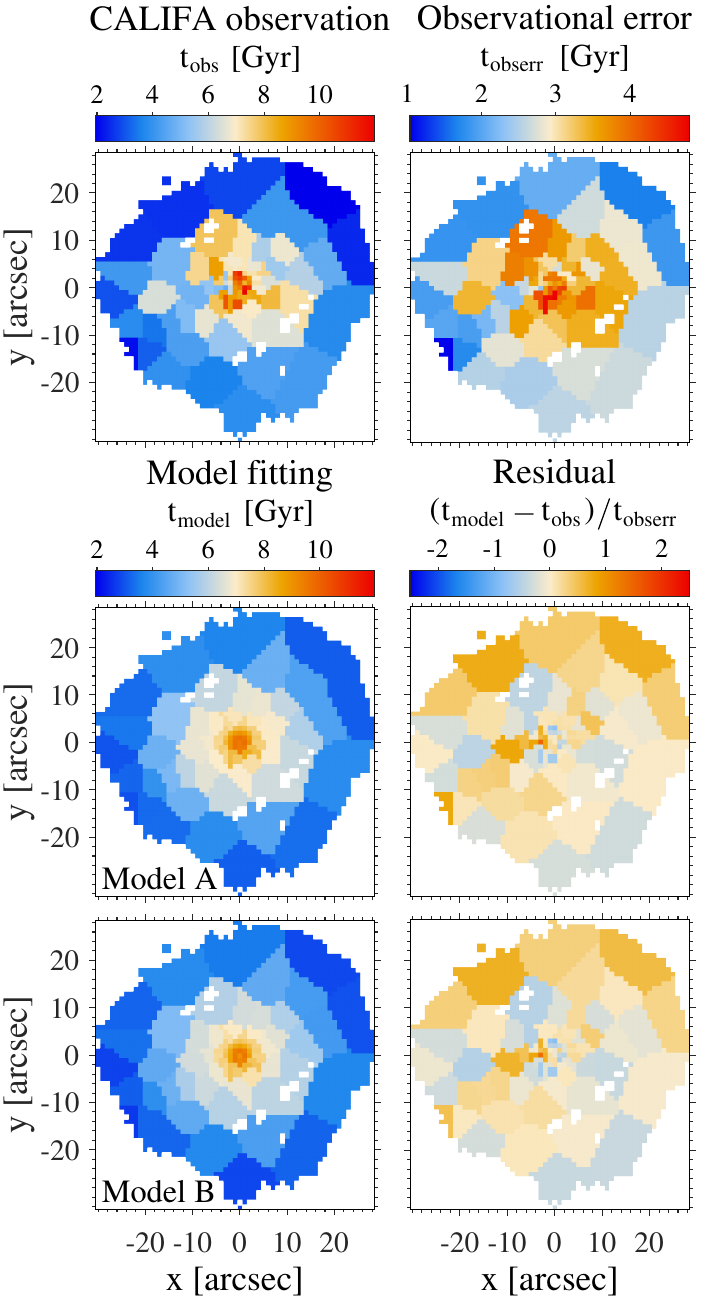}
    \caption{{\bf Similar to Fig.~\ref{fig8}, but using the new age error map.}}
    \label{figB1}
\end{centering}
\end{figure}
\begin{figure}
\begin{centering}
    \includegraphics[width=8.5cm]{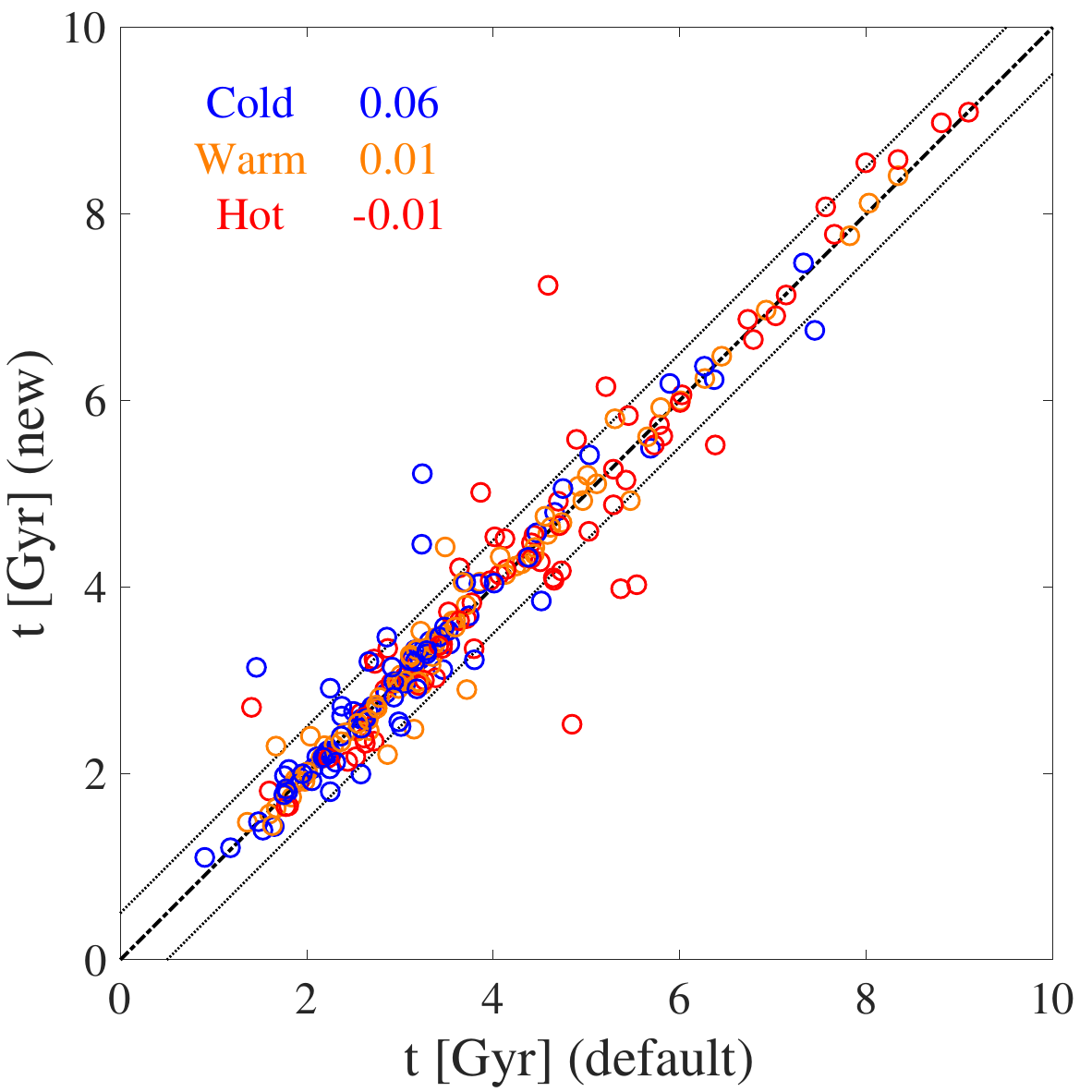}
    \caption{{\bf One-to-one comparison of the stellar ages of the cold, warm, and hot components derived from model B under different age error maps for 82 CALIFA spirals.} The $x$-axis represents ages calculated under the default error maps, as defined in Equation~\ref{age-err}, while $y$-axis represents ages calculated under the new error maps, as defined in Equation~\ref{age-err-new}. The blue, orange, and red circles denote cold, warm, and hot components respectively, and the corresponding colour texts show the mean difference of the ages of different components. The dashed line represent the equality of ages, while the dotted lines are $\pm0.5$ Gyr away from the dashed line.}
    \label{figB2}
\end{centering}
\end{figure}
\begin{figure}
\begin{centering}
    \includegraphics[width=8.5cm]{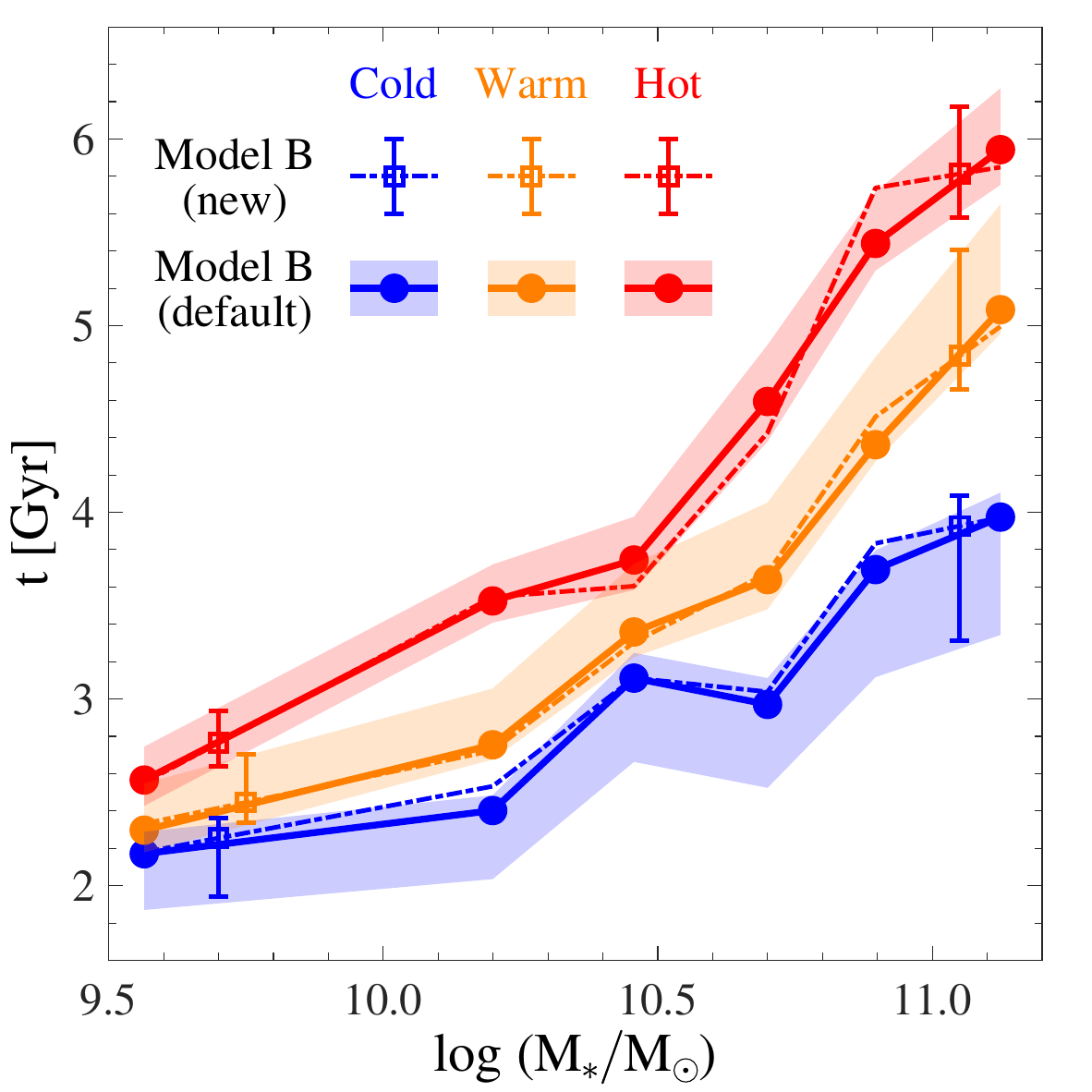}
    \caption{{\bf Similar to the bottom panel of Fig.~\ref{fig10}, but changing model A to model B with new age error maps.}}
    \label{figB3}
\end{centering}
\end{figure}

Here, we define a new version of age error maps. We still binned the age map in the same way as described in $\S$~\ref{sec4.3}, but define the age error in each bin by adapting the luminosity-weighted mean error of all pixels in that bin,
\begin{equation}
    t_{\rm obserr}^i=\frac{\sum_{j=1}^{N_i} L_j \sigma(t_j)}{\sum_{j=1}^{N_i} L_j}.\\
\label{age-err-new}
\end{equation}
In this way, the errors of the binned ages might be overestimated, especially in the outer regions of galaxies, in contrast to the errors defined in Equation~\ref{age-err}. We show this new version of the error map together with the corresponding best-fitting models for NGC 0171 in Fig.~\ref{figB1}. We note that we kept the results from model A unchanged as they are not influenced by the error map.\\

Using the new age error maps, we reran the models for all our sample galaxies with other settings unchanged and compared the recovered ages of the cold, warm, and hot components with previous results in Fig.~\ref{figB2}. We then replot the bottom panel of Fig.~\ref{fig10} by changing model A to model B with new age error maps, and present them in Fig.~\ref{figB3}. For different versions of error maps, the results from model B are statistically similar, which confirms the validity of our major conclusions.\\

\section{Data table}
\label{appendixC}
The main properties of 82 spiral galaxies in our sample and the corresponding modelling results are available in Table~\ref{data-table}.\\

\onecolumn
\begin{longtable}{|ccc|ccc|ccc|}
\caption{The main properties of 82 spiral galaxies in our sample and the corresponding modelling results.} \label{data-table}\\
\hline
\multicolumn{3}{|c|}{Galaxy properties} & \multicolumn{3}{c|}{Model A results} & \multicolumn{3}{c|}{Model B results}\\
\hline
Name & $\log(M_*/\rm M_{\odot})$ & $R_{\rm e}$ (kpc) & $t_{\rm cold}$ (Gyr) & $t_{\rm warm}$ (Gyr) & $t_{\rm hot}$ (Gyr) & $t_{\rm cold}$ (Gyr) & $t_{\rm warm}$ (Gyr) & $t_{\rm hot}$ (Gyr)\\
\hline
\endfirsthead
\hline
\multicolumn{3}{|c|}{Galaxy properties} & \multicolumn{3}{c|}{Model A results} & \multicolumn{3}{c|}{Model B results}\\
\hline
Name & $\log(M_*/\rm M_{\odot})$ & $R_{\rm e}$ (kpc) & $t_{\rm cold}$ (Gyr) & $t_{\rm warm}$ (Gyr) & $t_{\rm hot}$ (Gyr) & $t_{\rm cold}$ (Gyr) & $t_{\rm warm}$ (Gyr) & $t_{\rm hot}$ (Gyr)\\
\hline
\endhead
\hline
\multicolumn{9}{r}{Continued on next page.}\\
\endfoot
\hline
\endlastfoot
UGC12054 & 8.99 & 2.18 & 1.20 & 1.83 & 1.68 & 0.91 & 1.36 & 2.63 \\
UGC08231 & 9.13 & 3.26 & 1.53 & 1.91 & 2.42 & 1.18 & 2.15 & 2.20 \\
NGC0216 & 9.28 & 2.23 & 1.50 & 1.66 & 1.37 & 1.65 & 1.84 & 1.60 \\
NGC0755 & 9.38 & 3.28 & 1.87 & 2.02 & 1.94 & 2.22 & 1.98 & 1.93 \\
NGC5682 & 9.40 & 4.24 & 2.38 & 2.09 & 2.14 & 2.16 & 1.60 & 2.84 \\
NGC2604 & 9.66 & 3.82 & 1.83 & 2.00 & 1.76 & 1.81 & 2.19 & 1.81 \\
NGC3381 & 9.68 & 2.76 & 2.04 & 1.82 & 2.68 & 1.48 & 1.77 & 3.10 \\
NGC4961 & 9.68 & 2.67 & 2.03 & 2.05 & 1.99 & 1.93 & 2.01 & 2.18 \\
UGC12857 & 9.75 & 3.30 & 2.53 & 2.25 & 3.09 & 3.46 & 2.04 & 3.00 \\
UGC12816 & 9.82 & 5.99 & 2.63 & 1.73 & 2.43 & 2.07 & 1.88 & 2.73 \\
IC1151 & 9.85 & 3.34 & 1.64 & 2.25 & 2.84 & 2.11 & 2.16 & 3.05 \\
NGC5205 & 9.86 & 2.33 & 4.16 & 4.52 & 3.41 & 4.66 & 4.31 & 3.13 \\
NGC5520 & 9.86 & 1.60 & 2.48 & 3.73 & 3.37 & 2.58 & 4.58 & 3.16 \\
UGC12864 & 10.05 & 8.85 & 2.46 & 2.46 & 2.27 & 1.79 & 1.79 & 2.58 \\
NGC2730 & 10.12 & 6.54 & 1.61 & 2.81 & 2.78 & 1.53 & 3.24 & 2.87 \\
IC1528 & 10.14 & 6.30 & 1.98 & 1.89 & 2.63 & 2.38 & 1.63 & 2.72 \\
NGC5480 & 10.14 & 3.37 & 1.67 & 1.86 & 2.03 & 1.76 & 1.67 & 2.53 \\
NGC6063 & 10.14 & 4.08 & 2.60 & 2.34 & 3.48 & 3.32 & 1.90 & 3.20 \\
UGC04280 & 10.14 & 2.70 & 2.20 & 3.57 & 3.90 & 2.50 & 3.14 & 3.77 \\
NGC6132 & 10.21 & 5.05 & 2.45 & 2.98 & 3.58 & 2.85 & 3.01 & 3.14 \\
UGC09476 & 10.21 & 4.93 & 2.53 & 3.33 & 4.05 & 2.66 & 3.42 & 3.71 \\
NGC5016 & 10.23 & 3.23 & 1.90 & 2.70 & 4.38 & 1.80 & 2.99 & 4.06 \\
UGC08778 & 10.24 & 3.47 & 3.55 & 3.68 & 4.01 & 3.25 & 2.97 & 5.42 \\
NGC5633 & 10.26 & 2.23 & 2.06 & 2.65 & 3.78 & 2.22 & 2.01 & 4.36 \\
NGC3687 & 10.27 & 2.95 & 3.14 & 4.65 & 4.52 & 3.06 & 4.62 & 4.44 \\
UGC10384 & 10.27 & 4.07 & 2.38 & 2.60 & 2.37 & 1.46 & 1.67 & 1.41 \\
NGC5657 & 10.28 & 2.90 & 2.27 & 2.79 & 3.49 & 2.25 & 3.15 & 3.38 \\
NGC4210 & 10.29 & 4.02 & 3.18 & 4.62 & 4.38 & 3.21 & 4.13 & 5.29 \\
NGC0237 & 10.31 & 4.45 & 2.65 & 2.82 & 2.73 & 2.58 & 2.87 & 2.73 \\
NGC3815 & 10.35 & 3.67 & 3.37 & 4.16 & 2.88 & 3.74 & 3.56 & 2.90 \\
MCG-02-02-030 & 10.37 & 4.74 & 3.73 & 3.36 & 4.33 & 4.52 & 3.35 & 3.64 \\
NGC6762 & 10.38 & 1.93 & 2.88 & 3.40 & 3.15 & 2.99 & 3.33 & 3.26 \\
NGC2906 & 10.39 & 2.88 & 3.35 & 3.74 & 3.86 & 3.16 & 3.47 & 4.50 \\
NGC3811 & 10.42 & 4.55 & 2.32 & 2.71 & 3.03 & 2.70 & 2.67 & 2.44 \\
NGC3994 & 10.42 & 2.03 & 3.37 & 3.35 & 2.99 & 3.11 & 3.71 & 3.46 \\
NGC4644 & 10.45 & 4.33 & 3.32 & 4.83 & 4.19 & 3.20 & 4.96 & 4.65 \\
NGC0477 & 10.50 & 8.85 & 2.35 & 2.33 & 2.37 & 2.37 & 2.54 & 2.62 \\
NGC7653 & 10.50 & 3.75 & 2.73 & 2.68 & 3.77 & 2.26 & 2.74 & 4.41 \\
NGC2540 & 10.52 & 6.20 & 1.95 & 2.32 & 2.71 & 2.19 & 2.41 & 2.62 \\
NGC7631 & 10.53 & 4.53 & 3.57 & 4.04 & 4.41 & 3.23 & 3.86 & 5.54 \\
NGC6310 & 10.56 & 5.61 & 5.53 & 4.84 & 4.71 & 4.47 & 4.74 & 6.02 \\
NGC6186 & 10.57 & 4.12 & 2.97 & 3.39 & 3.92 & 2.62 & 3.38 & 4.64 \\
UGC09067 & 10.58 & 7.96 & 3.17 & 3.06 & 3.29 & 3.53 & 2.78 & 2.72 \\
NGC7549 & 10.60 & 6.60 & 2.30 & 3.15 & 2.66 & 2.25 & 1.98 & 1.78 \\
NGC0551 & 10.64 & 7.13 & 4.87 & 3.07 & 4.53 & 5.03 & 3.05 & 5.03 \\
NGC0234 & 10.65 & 6.24 & 2.07 & 2.32 & 3.52 & 2.21 & 2.29 & 3.45 \\
MCG-02-51-004 & 10.68 & 6.97 & 3.08 & 2.85 & 5.34 & 2.93 & 2.97 & 5.78 \\
NGC0776 & 10.69 & 6.47 & 4.07 & 4.20 & 4.34 & 2.91 & 3.37 & 5.82 \\
NGC6004 & 10.69 & 5.89 & 2.47 & 3.12 & 2.44 & 2.37 & 3.38 & 2.23 \\
NGC0171 & 10.72 & 7.22 & 4.35 & 5.77 & 9.15 & 3.85 & 6.01 & 9.10 \\
NGC5000 & 10.73 & 6.29 & 1.97 & 5.18 & 4.59 & 2.06 & 4.91 & 4.70 \\
UGC12810 & 10.73 & 11.53 & 2.42 & 3.46 & 4.76 & 2.64 & 3.60 & 4.71 \\
NGC2916 & 10.75 & 6.75 & 3.32 & 4.43 & 3.90 & 3.52 & 4.45 & 3.64 \\
NGC7466 & 10.75 & 7.01 & 2.11 & 3.23 & 3.36 & 2.17 & 3.11 & 3.52 \\
NGC2487 & 10.77 & 9.42 & 3.90 & 5.01 & 4.25 & 3.70 & 4.55 & 5.37 \\
NGC0001 & 10.80 & 3.99 & 2.69 & 3.43 & 3.67 & 2.86 & 3.18 & 3.79 \\
UGC10388 & 10.82 & 3.55 & 5.79 & 6.05 & 7.70 & 6.37 & 6.27 & 7.66 \\
UGC00005 & 10.83 & 8.41 & 3.31 & 2.71 & 5.23 & 3.48 & 2.66 & 6.01 \\
NGC0192 & 10.85 & 6.46 & 1.55 & 2.53 & 3.49 & 1.75 & 2.37 & 3.97 \\
NGC7364 & 10.88 & 4.26 & 2.79 & 4.08 & 4.58 & 2.59 & 3.20 & 5.45 \\
NGC0036 & 10.90 & 9.19 & 3.26 & 4.19 & 4.19 & 3.16 & 4.44 & 4.41 \\
NGC6394 & 10.90 & 8.65 & 4.13 & 4.25 & 6.25 & 2.67 & 3.49 & 4.59 \\
IC0674 & 10.91 & 5.15 & 6.72 & 8.27 & 8.79 & 6.26 & 8.34 & 8.81 \\
NGC0180 & 10.92 & 10.45 & 5.44 & 4.66 & 4.39 & 5.89 & 4.25 & 4.73 \\
IC1755 & 10.93 & 6.30 & 6.00 & 7.48 & 7.08 & 5.69 & 6.93 & 7.57 \\
NGC6060 & 10.93 & 8.92 & 2.25 & 3.19 & 4.26 & 2.31 & 3.11 & 4.13 \\
NGC7321 & 10.93 & 7.42 & 2.72 & 4.30 & 4.52 & 2.93 & 3.68 & 4.02 \\
NGC2347 & 10.94 & 5.74 & 3.15 & 2.83 & 5.17 & 3.01 & 3.22 & 4.89 \\
IC4566 & 10.95 & 6.12 & 3.45 & 5.73 & 5.45 & 3.18 & 5.01 & 5.21 \\
NGC7563 & 10.96 & 2.95 & 4.73 & 5.57 & 6.19 & 3.24 & 5.31 & 6.38 \\
NGC6478 & 11.01 & 11.00 & 3.23 & 2.71 & 3.62 & 3.29 & 2.55 & 4.14 \\
NGC6301 & 11.02 & 14.24 & 3.57 & 2.33 & 4.92 & 3.42 & 2.76 & 6.79 \\
NGC0160 & 11.03 & 8.07 & 8.01 & 7.80 & 6.69 & 7.33 & 7.82 & 6.73 \\
NGC6497 & 11.04 & 2.85 & 4.83 & 5.75 & 5.64 & 4.01 & 5.11 & 5.73 \\
NGC7311 & 11.07 & 3.96 & 3.98 & 4.07 & 4.17 & 3.80 & 4.07 & 3.87 \\
NGC7738 & 11.08 & 6.95 & 3.43 & 4.18 & 4.19 & 2.25 & 3.72 & 4.84 \\
NGC2639 & 11.17 & 3.93 & 2.36 & 3.74 & 2.81 & 1.78 & 3.59 & 3.42 \\
UGC06036 & 11.17 & 5.39 & 8.94 & 8.22 & 7.72 & 7.45 & 8.03 & 8.00 \\
NGC3106 & 11.21 & 9.20 & 6.98 & 6.99 & 6.00 & 4.75 & 5.80 & 7.03 \\
NGC5888 & 11.21 & 9.80 & 4.35 & 6.62 & 8.37 & 4.38 & 6.45 & 8.34 \\
NGC5987 & 11.21 & 6.91 & 2.17 & 5.71 & 7.03 & 1.95 & 5.66 & 7.14 \\
NGC5406 & 11.27 & 7.55 & 3.46 & 5.59 & 5.31 & 3.29 & 5.47 & 5.29
\end{longtable}
\tablefoot{From left to right, they are (1) galaxy name; (2) stellar mass $\log(M_*/\rm M_{\odot})$; (3) the effective radius $R_{\rm e}$; (4) the mean age of the cold component derived from model A; (5) the mean age of the warm component derived from model A; (6) the mean age of the hot component derived from model A; (7) the mean age of the cold component derived from model B; (8) the mean age of the warm component derived from model B; (9) the mean age of the hot component derived from model B. We refer to the journal website for the complete table.}

\end{appendix}

\end{nolinenumbers}
\end{document}